\documentclass[11pt]{article}
\usepackage{amssymb}
\usepackage{graphics}
\usepackage{epsfig}
\usepackage{a4wide}
\usepackage{color}

\begin{document}

\title{ NNLO QCD corrections to Higgs pair production via vector boson fusion at hadron colliders }
\author{ Ling Liu-Sheng, Zhang Ren-You, Ma Wen-Gan, Guo Lei, Li Wei-Hua, and Li Xiao-Zhou\\
{\small Department of Modern Physics, University of Science and Technology of China (USTC),} \\
{\small Hefei, Anhui 230026, People's Republic of China}}
\date{}
\maketitle \vskip 15mm

\begin{abstract}
The measurement of the Higgs pair production via vector boson fusion
can be used to test the trilinear Higgs self-coupling and the $VVHH$
$(V=Z,W)$ quartic gauge interactions. In this paper we present the
calculations of the next-to-next-to-leading-order QCD
corrections to the SM Higgs boson pair production via vector boson
fusion at hadron colliders with the center-of-mass energy of
$14$, $33$, and $100~{\rm TeV}$ by using the
structure function approach, and study the residual uncertainties
from the factorization/renormalization scale, parton distribution
functions and $\alpha_s$ on the total cross section. We also provide
the distributions of transverse momenta, rapidities, invariant mass
and azimuthal angle separations of final Higgs bosons. We observe a
considerable quantitative reduction in the scale uncertainty due to
the next-to-next-to-leading-order QCD corrections, and find that the
total cross section is sensitive to the trilinear Higgs self-coupling.
\end{abstract}

\vskip 3cm


\vfill \eject
\baselineskip=0.32in

\vskip 5mm
\section{Introduction}
\par
In the Standard Model (SM) and its extensions, the Higgs boson is
responsible for the electroweak symmetry breaking (EWSB) and the
generation of elementary particle masses. One of the primary goals
of the LHC is to uncover the origin of EWSB
and to determine whether a SM Higgs boson exists. A giant step was
made recently; both the ATLAS and CMS collaborations have observed a
new boson with the mass of $\sim 126~{\rm GeV}$, and its properties
are, so far, compatible with the SM Higgs \cite{1-ATCM}. The next
important step is to investigate whether this particle is indeed
responsible for the EWSB and, eventually, to determine whether it is
really the SM Higgs boson. To do so, it is crucial to probe the
Higgs self-interactions, since they trigger the EWSB and are
indispensable to reconstruct the Higgs potential
\cite{2-self,3-self,4-self}.

\par
The Higgs pair production at hadron colliders is sensitive to the
trilinear Higgs self-coupling. There are four main Higgs pair production
channels: gluon-gluon fusion via top-quark loop, vector boson fusion (VBF),
top-quark pair associated production and double Higgs-strahlung
\cite{6-triple}. Among these Higgs pair production mechanisms, the
gluon-gluon fusion mechanism provides the largest cross section,
while the VBF mechanism yields the second largest cross section,
which is quantitatively 1 order smaller than that via the former
one. The VBF mechanism shows a clear experimental signature of two
centrally produced Higgs bosons and two highly energetic
forward/backward jets \cite{9-vbfsig,10-vbfsig}, but the event
analysis is still challenged by the smallness of its cross section
\cite{6-triple,7-triple}.
Therefore, a study of the VBF Higgs pair production can be
feasible only at high luminosity and very high energy hadron
colliders \cite{8-HLHC,8-VLHC}. At these hadron colliders, the Higgs
pair production via weak vector boson fusion is not only the leading
process, which is sensitive to the $W^+W^-HH$ and $ZZHH$
interactions but also can be used to study the EWSB by probing
trilinear Higgs self-coupling. In Ref.\cite{11-struf} Paolo Bolzoni
{\it et al.} pointed out that the structure function approach
\cite{12-struf} and the QCD factorization approximation work
extremely well up to ${\cal O}(\alpha_s^2)$ corrections for the VBF
processes, and the remaining contributions which are kinematically
and parametrically suppressed, are practically negligible.
The next-to-leading-order (NLO) and next-to-next-to-leading-order (NNLO)
QCD corrections to the VBF single Higgs production at the
LHC have been evaluated by using the structure function approach in
Refs. \cite{11-struf} and \cite{12-struf}, separately.

\par
In this work we present the calculations of the VBF Higgs pair
production at hadron colliders with high luminosity or very high
energy up to the QCD NNLO by using the structure function approach.
The paper is organized as follows. In Sec. 2, we give a brief
description of the structure function approach, and the strategy of
the QCD NNLO calculation. The numerical results and discussion are
presented in Sec. 3. A short summary is given in Sec. 4. In the Appendix
the explicit expressions for coefficients $C_{ij}~(i,j = 1,2,3)$ are
provided.

\vskip 5mm
\section{Calculation setup}
\par
\begin{figure*}
\begin{center}
\includegraphics[scale=0.8]{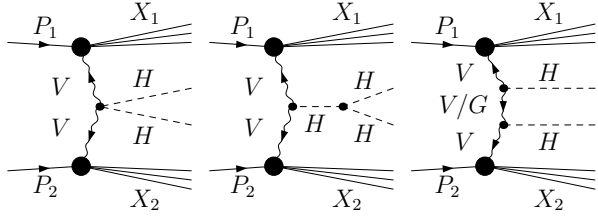}
\caption{\label{fig1} VBF Higgs pair production process at the hadron
collider. }
\end{center}
\end{figure*}

The structure function approach is a very good approximation to the
VBF processes at hadron colliders, which is accurate at a precision
level well above the typical residual scale and parton distribution
function (PDF) uncertainties \cite{11-struf}. This approximation is
based on the absence or
smallness of the QCD interference between the two inclusive final
proton remnants. The mechanism of the VBF Higgs pair production is
analogous to the VBF single Higgs production. It can be viewed as
the double deep-inelastic scattering (DIS) of two (anti)quarks with
two virtual weak vector bosons independently emitted from the
hadronic initial states fusing into a Higgs boson pair
\cite{7-triple}. In particular, the interference between the Higgs
pair radiated off the fusing weak vector bosons and the double
Higgs-strahlung process via $qq' \to HHV^* \to HHqq'$ is negligible,
and therefore the latter process is treated separately. Furthermore,
the VBF Higgs pair production event can be easily selected because
it includes two widely separated jets with high invariant mass.
Therefore, we can use the structure function approach to provide the
precision predictions at the QCD NNLO accuracy for the VBF Higgs
pair production process at hadron colliders as used in the
calculations for the VBF single Higgs production. The Feynman
diagrams for the VBF Higgs pair production in proton-proton
collisions are depicted in Fig.\ref{fig1}, where $P_i~(i=1,2)$
denote the 4-momenta of the initial protons, the virtual vector
boson $V$ can be either $W$ or $Z$, $G$ stands for the Goldstone
boson, and $X_i~(i=1,2)$ are the proton remnants.
\par
\begin{figure*}
\begin{center}
\includegraphics[scale=1]{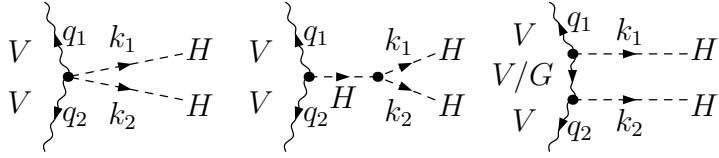}
\caption{\label{fig2} The Feynman diagrams for the $VV \to HH$ process. }
\end{center}
\end{figure*}

By applying the structure function approach, the cross section for the
VBF Higgs pair production can be calculated by contracting the DIS
hadronic tensor $W_{\mu \nu}$ with the matrix element of the vector
boson fusion subprocess ${\cal{M}}^{\mu \nu}_V$. The leading-order
(LO) Feynman graphs for the $VV \to HH$ process are shown in
Fig.\ref{fig2}. The differential cross section for the VBF Higgs
pair production process can be expressed as
\begin{equation}\label{factorization}
d\sigma = \sum_{V=Z,W} d\sigma_V,                                                                   \nonumber \\
\end{equation}
where
\begin{eqnarray}\label{factorization-1}
d\sigma_V &=& \frac{G_F^2 M^4_V }{S (Q^2_1 +M^2_V)^2 (Q^2_2 +M^2_V)^2}
W_{\mu\nu}(x_1, Q^2_1) {\cal {M}}^{\mu\rho}_V {\cal{M}}^{*\nu\sigma}_V W_{\rho\sigma}(x_2,Q^2_2)   \nonumber \\
&&\times
\frac{d^3 P_{X_1}}{(2\pi)^3 2 E_{X_1}}  \frac{d^3 P_{X_2}}{(2\pi)^3 2 E_{X_2}} ds_1 ds_2
\frac{d^3 k_1}{(2\pi)^3 2 E_1} \frac{d^3 k_2}{(2\pi)^3 2 E_2}                                      \nonumber \\
&&\times
(2\pi)^4\delta^4( P_1+P_2-P_{X_1} -P_{X_2}-k_1 - k_2 )\,.
\end{eqnarray}
Here $d\sigma_V$ stands for the contribution of the $VV$ $(V=Z,W)$
fusion process, $G_F$ denotes the Fermi constant, $S$ is the
proton-proton colliding energy squared in the center-of-mass system
(c.m.s), $M_{V}$ is the mass of vector boson $V$, ${Q_i}^2 =
-q_{i}^2$, and $x_i={Q_i}^2/(2 P_i \cdot q_i)$ are the usual DIS
variables, and $s_i=(P_i+q_i)^2$ is the invariant mass squared of
the $i$ th proton remnant. By adopting the Feynman gauge, the matrix
element of $VV$ fusion subprocess can be expressed as
\begin{eqnarray}\label{matrix}
{\cal{M}}^{\mu\nu}_V &=& 2 \sqrt{2} G_F g^{\mu\nu} \left[\frac{2
M^4_V}{(q_1+k_1)^2-M^2_V}
 +\frac{2 M^4_V}{(q_1+k_2)^2-M^2_V}+\frac{6 v\lambda^{SM}_{HHH} M^2_V}{(k_1+k_2)^2-M^2_H}+M^2_V\right]      \nonumber  \\
&&+ \frac{\sqrt{2} G_F M^2_V}{(q_1+k_1)^2-M^2_V}(2 k^\mu_1+q^\mu_1)(-k^\nu_1+k^\nu_2-q^\nu_1)     \nonumber\\
&&+ \frac{\sqrt{2} G_F M^2_V}{(q_1+k_2)^2-M^2_V}(2
k^\mu_2+q^\mu_1)(k^\nu_1-k^\nu_2-q^\nu_1),     \,
\end{eqnarray}
where $M_V$ is the mass of $W$ or $Z$, $\lambda^{SM}_{HHH} =
\frac{M^2_H}{2v}$ is the SM trilinear Higgs self-coupling, and $v$ is
the vacuum expectation value of Higgs field. The DIS hadronic tensor
has the form as \cite{14-PDG}
\begin{eqnarray}
W_{\mu \nu}\left(x_i, Q_i^2\right) &=&
\left( - g_{\mu \nu} + \frac{q_{i\mu} q_{i\nu}}{q_i^2} \right)\, F_{1}(x_i,Q_i^2)
 + \frac{ {\hat P}_{i\mu}{\hat P}_{i\nu} }{P_i \cdot q_i}\, F_{2}(x_i,Q_i^2) \nonumber \\
 &+& {\rm{i}} \epsilon_{\mu\nu\alpha\beta} \frac{P_i^\alpha q_i^\beta}{2 P_i \cdot q_i} F_{3}(x_i,Q_i^2), ~~~~~(i=1,2),
\end{eqnarray}
where ${\hat P}_{i} = P_{i} - \frac{P_i \cdot q_i}{q_i^2}\, q_{i}$
and $F_j\left(x_i, Q_i^2\right)(j = 1,2,3)$ are the usual DIS
structure functions of proton \cite{20-struf}.

\par
For the VBF Higgs pair production the interferences between the $u$ and
$t$ channels with identical final quarks (e.g., $uu \to HHuu$), and
between the processes with $WW$ and $ZZ$ fusions (e.g., $ud \to
ZZ/WW \to HHud$) at the LO, NLO, and NNLO in QCD are normally
nonfactorizable. These nonfactorizable contributions would make
Eq.(\ref{factorization}) being incorrect even at the LO. However, these
interference effects are heavily suppressed by kinematics for the
VBF Higgs pair production. We have calculated these interference
contributions at the LO by applying FeynArts-3.7 and FormCalc-7.4
packages \cite{21-feyn} and found that they contribute less than
$0.01\%$ to the total cross section. Therefore, it is reasonable to
neglect these interference contributions in the QCD LO, NLO, and NNLO
calculations. Apart from these interference effects, in the QCD NNLO
calculation, the diagrams involving the exchange of gluon between the
two quark lines are also nonfactorizable. The same as in the VBF single
Higgs production case \cite{13-struf}, this nonfactorizable
correction at the QCD NNLO is negligible for the VBF Higgs pair
production.

\par
We express the matrix element squared as
\begin{eqnarray}\label{Cij}
W_{\mu\nu}\left(x_1, Q^2_1\right) M^{\mu\rho} {M^*}^{\nu\sigma}
W_{\rho\sigma}\left(x_2,Q^2_2\right) &=& \sum_{i,j=1}^3 F_i
\left(x_1, Q^2_1\right) F_j \left(x_2,Q^2_2\right) C_{ij}.
\end{eqnarray}
The explicit expressions for $C_{ij}~(i,j = 1,2,3)$ are collected in the
Appendix. The explicit expressions for the DIS structure functions
at the LO and NLO have been given in Refs.\cite{9-vbfsig,15-FF}, and
the NNLO expressions can be found in Refs.\cite{11-struf,13-struf}.
In general, the DIS structure functions are expressed as
convolutions of the PDFs with the
Wilson coefficient functions $C_i~(i = 1,2,3)$. There are a number
of PDFs at the QCD NNLO accuracy available, e.g., ABM11
\cite{17-abm}, CT10 \cite{22-ct}, HERAPDF1.5 \cite{23-hera},
MSTW2008 \cite{24-mstw}, and NNPDF2.3 \cite{21-nnpdf}. The Wilson
coefficients can be obtained up to the QCD NNLO
from Refs.\cite{16-CC,17-CC,18-CC}, and the accurate parametrization
of them can be taken from Ref.\cite{19-CC}. We developed a Fortran
program to evaluate the numerical results for the VBF Higgs pair
production process by employing the structure function approach.
To verify the correctness of our calculations, we use our
Fortran code to calculate the VBF Higgs pair production process at
the QCD NLO accuracy by taking the same conditions as in
Ref.\cite{7-triple}, i.e., adopting the structure function approach
and the MSTW2008 ($90\%$ C.L.) PDFs, setting $\mu=\mu_f = \mu_r = Q$,
$M_H = 125~{\rm GeV}$, and the other parameters being also the same
as in Ref.\cite{7-triple}. Our numerical results of the total cross
section are in good agreement with those in Table 3 of Ref.\cite{7-triple}
implemented in the VBFNLO code \cite{27-vbfnlo}; e.g., we
get $\sigma_{qq'HH}^{NLO}=2.009(1)~fb$ at the $\sqrt{S}=14~{\rm
TeV}$ LHC, which is coincident with $\sigma_{qq'HH}^{NLO}=2.01~fb$ in
Ref.\cite{7-triple}.

\vskip 5mm
\section{Numerical results and discussion }
\par
In this section we present and discuss the numerical results with
the corrections up to the NNLO in QCD to the VBF Higgs pair production
at the $\sqrt{S}=14$, $33$, and $100~{\rm TeV}$ proton-proton colliders.
In further numerical calculations, we mainly use the MSTW2008 ($68\%$
C.L.) PDFs \cite{24-mstw} with the default value of strong coupling
constant required by the set, while in comparison of the results by
adopting different PDFs, we use separately the ABM11, CT10,
HERAPDF1.5, MSTW2008 (68$\%$ C.L.) and NNPDF2.3 PDFs. The related SM
input parameters are taken as \cite{14-PDG}
\begin{eqnarray}
M_H = 126~{\rm GeV},~M_W = 91.1876~{\rm GeV},~~M_Z = 80.385~{\rm
GeV},~~G_F = 1.1663787 \times 10^{-5}~{\rm GeV}^{-2}.
\end{eqnarray}
A cut of $Q_i^2 > 4~{\rm GeV}^2$ has been applied in order to render
the results in the perturbative regime.

\par
\subsection{Cross sections and uncertainties}
\par
To make a strict cross section comparison between the
theoretical predictions and experimental results, we should assess
thoroughly the uncertainties affecting the central predictions of
the total cross sections. In this section, we will discuss three
kinds of uncertainties: (1) the scale uncertainty, which is an
estimate of the missing higher-order contributions in the
perturbative calculation; (2) the PDF uncertainty; and (3) the
uncertainty related to the fitted value of the strong coupling
constant $\alpha_s(M_Z^2)$ and the parametric uncertainties related
to the experimental errors.

\par
\subsubsection{Cross sections and scale uncertainty}
\par
\begin{figure*}
\begin{center}
\includegraphics[scale=0.65]{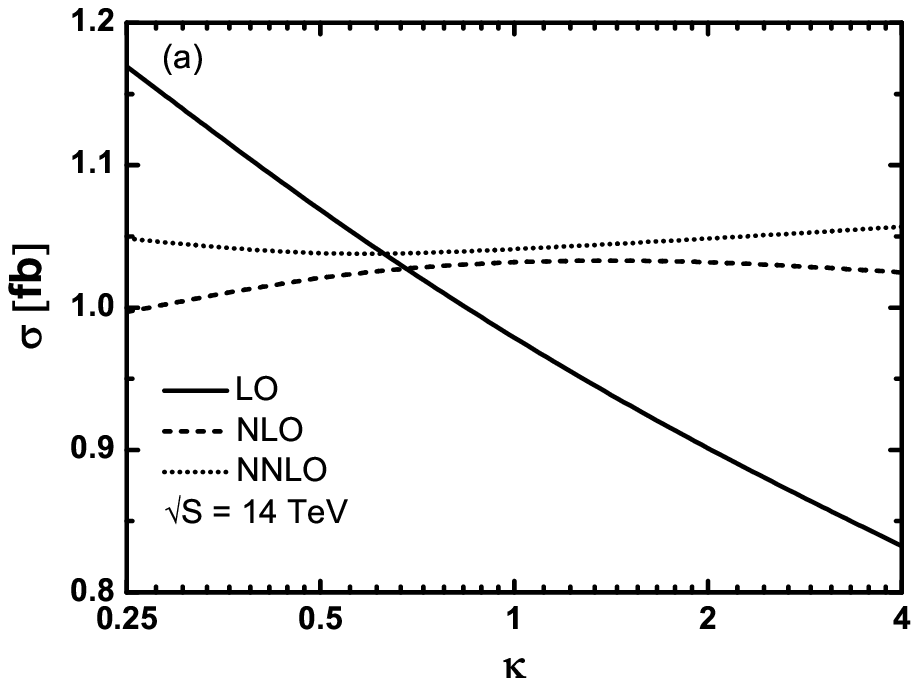}
\includegraphics[scale=0.65]{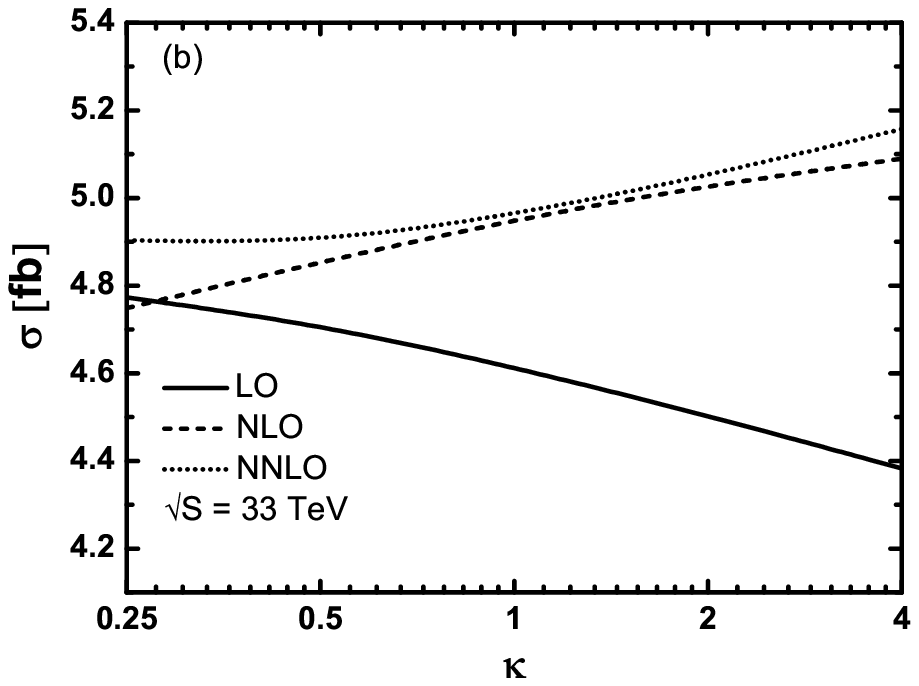}
\includegraphics[scale=0.65]{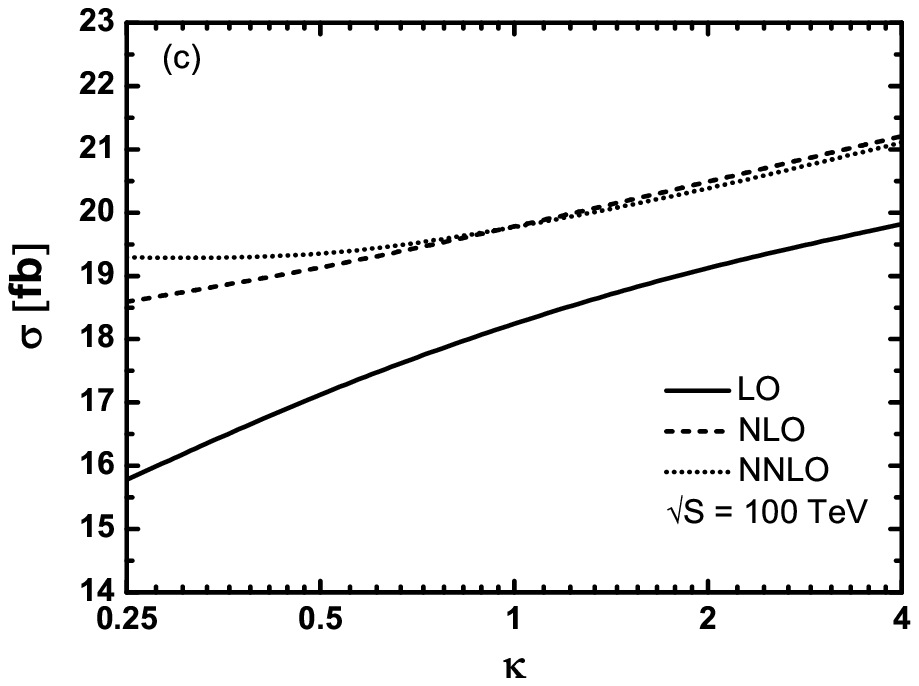}
\caption{\label{fig3} The scale dependence of the total cross
section with $\mu=\kappa Q$ and $\kappa \in [1/4, 4]$ by using
the MSTW2008 ($68\%$ C.L.) PDFs. (a) $\sqrt{S}=14~{\rm TeV}$.
(b) $\sqrt{S}=33~{\rm TeV}$. (c)$\sqrt{S}=100~{\rm TeV}$.}
\end{center}
\end{figure*}

The theoretical prediction of the cross section depends on the
factorization scale $\mu_f$, which originates from the convolution of
the perturbative partonic cross section with the nonperturbative
PDFs, and the renormalization scale $\mu_r$ that comes from the
running of $\alpha_s$. An estimate of the missing higher-order
corrections can be considered as the variation of the central cross
section with respect to these two scales. For simplicity we take the
factorization scale being equal to the renormalization scale, i.e.,
$\mu=\mu_f=\mu_r$, and define $\mu=\kappa \mu_0$. We fix the central
scale value as $\mu_0=Q$ with $\mu$ varying in the range of $[0.25
\mu_0,~ 4 \mu_0]$. There, the central scale is the virtuality of the
vector bosons which fuse into the Higgs boson pair. That is the most
natural central scale choice for VBF processes \cite{11-struf}. In
Figs.\ref{fig3}(a,b,c), we depict the scale dependence of the
integrated cross section for the VBF Higgs pair production process
at the $\sqrt{S}=14$, $33$, and $100~{\rm TeV}$
hadron colliders, separately, by using the MSTW2008 ($68\%$ C.L.)
PDFs. If we define the scale uncertainty quantitatively as
\begin{eqnarray}
\zeta \equiv
\frac{max[\sigma(\mu)]-min[\sigma(\mu)]}{\sigma(\mu_0)},~~
(\mu \in [1/4\mu_0, 4\mu_0]),
\end{eqnarray}
the scale uncertainty parameter $\zeta$ at the $14~{\rm TeV}$ LHC
can reach the value of $35\%$ at the LO and is reduced to $3.9\%$ by
the NLO QCD corrections, while the NNLO scale uncertainty decreases
to $2.3\%$. The LO, NLO, and NNLO scale uncertainties have the values
of $15\%$, $6.3\%$, and $3.5\%$ at the $33~{\rm TeV}$ hadron collider
and the values of $15\%$, $11.5\%$, and $5.9\%$ at the $100~{\rm TeV}$
hadron collider, respectively. We can see that the value of $\zeta$
at the QCD NNLO accuracy is less than the scale uncertainty of the
NLO QCD corrected cross section at these hadron colliders.

\par
\begin{figure*}
\begin{center}
\includegraphics[scale=0.65]{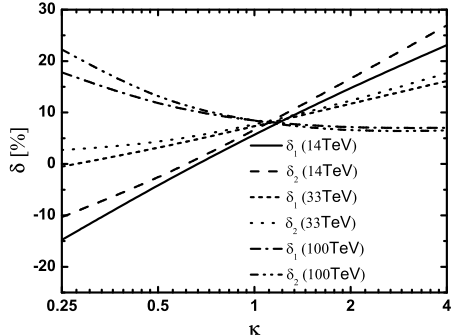}
\caption{\label{fig4} The relative QCD corrections, $\delta_1$ and
$\delta_2$, as functions of $\kappa$ at the $\sqrt{S}=14$,
$33$, and $100~{\rm TeV}$ hadron colliders with $\kappa \in
[1/4, 4]$. }
\end{center}
\end{figure*}

Figure \ref{fig4} shows the relative QCD corrections, $\delta_1$ and
$\delta_2$, as functions of $\kappa$ at the $14$, $33$, and
$100~{\rm TeV}$ hadron colliders by using the MSTW2008
($68\%$ C.L.) PDFs, where we define $\delta_1
=\frac{(\sigma_{NLO}-\sigma_{LO})}{\sigma_{LO}}$ and $\delta_2
=\frac{(\sigma_{NNLO}-\sigma_{LO})}{\sigma_{LO}}$ to describe the
relative NLO and NNLO QCD corrections separately. In Fig.\ref{fig4},
the relative QCD corrections are obviously dependent on the value of
$\kappa$; particularly, the curves for $\delta_1$ are more
intensively related to the scale than the corresponding $\delta_2$
curves. We find that at the $\sqrt{S}=14~{\rm TeV}$ LHC $\delta_1$
and $\delta_2$ are $-15\%$ and $-9.0\%$ at the position of
$\kappa=0.25$ but change to be $23\%$ and $27\%$ separately when
$\kappa=4$. Analogous to Fig. \ref{fig3}, the results concerning the
relative corrections at the NNLO accuracy are less related to the
renormalization and factorization scales than those at the NLO
accuracy. That means the NNLO QCD corrections are very important in
improving the scale uncertainty and make it possible to take full
advantage of modern PDF sets at the same accuracy. We can also see
from Fig.\ref{fig4} that at the position of the central scale
($\kappa=1$) the relative corrections $\delta_1$ and $\delta_2$ are
$6.4\%$ and $6.9\%$ at the $\sqrt{S}=14~{\rm TeV}$ LHC and have the
values of $6.8\%~(5.9\%)$ and $7.2\%~(6.2\%)$ at the
$\sqrt{S}=33~(100)~{\rm TeV}$ hadron collider, respectively. These
results show also the choice of $\kappa=1$ can keep better
convergence in the perturbative calculation than other values of
$\kappa$. Therefore, we fix the scale $\mu=\mu_0$ in the following
calculations unless stated otherwise.

\par
In Table \ref{tab1}, we list the central values of the total cross
section ($\kappa=1$) and the errors due to scale uncertainty with
$\kappa$ varying in the range of $[1/4, 4]$ for the VBF Higgs pair
production process at the LO, NLO, and NNLO  by using the MSTW2008
($68\%$ C.L.) PDFs. We can read from the table that the NNLO QCD
corrected total cross section goes up from
$1.986^{+0.045}_{-0}$ fb to $80.05^{+3.92}_{-0.80}$ fb as the
increment of the hadron collider c.m.s. colliding energy $\sqrt{S}$
from $14$ to $100~{\rm TeV}$, and the scale uncertainty of
$\sigma_{NNLO}$ is much smaller than the corresponding ones of
$\sigma_{LO}$ and $\sigma_{NLO}$. Therefore, we can see that from
the point of view of improving the scale uncertainty the cross
section prediction including the NNLO QCD corrections is more
helpful for precision measurement of the VBF Higgs pair production
process.
\begin{table}
\begin{center}
\begin{tabular}{|c|c|c|c|}
\hline $\sqrt{S}$  &LO [fb]                    &NLO [fb]                   &NNLO [fb]                   \\
\hline   14 TeV    &$1.858^{+0.374}_{-0.270}$  &$1.976^{+0}_{-0.078}$      &$1.986^{+0.045}_{-0}$       \\
         33 TeV    &$11.234^{+0.878}_{-0.830}$ &$12.002^{+0.190}_{-0.562}$ &$12.041^{+0.359}_{-0.060}$  \\
        100 TeV    &$75.36^{+4.91}_{-6.34}$    &$79.82^{+3.92}_{-5.26}$    &$80.05^{+3.92}_{-0.80}$     \\
\hline
\end{tabular}
\caption{\label{tab1} The central values of the total cross section
($\kappa=1$) and the errors due to scale uncertainty with $\kappa
\in [1/4,4]$ at the $\sqrt{S}=14$, $33$, and $100~{\rm TeV}$ hadron
colliders by using the MSTW2008 ($68\%$ C.L.)
PDFs. }
\end{center}
\end{table}

\par
\subsubsection{PDF and $\alpha_s$ uncertainties}
\par
Except the theoretical scale uncertainty, there is another source of
theoretical uncertainty which is from the assumptions made on the
parametrization of the PDFs. It is a pure theoretical error due to
the parametrization choice, the set of input parameters used, the
running of the parameters, etc. One way to quantify the pure
theoretical uncertainties induced by these differences is to compare
the predictions obtained with the various PDF sets, such as the
ABM11 \cite{17-abm}, CT10 \cite{22-ct}, HERAPDF1.5 \cite{23-hera},
MSTW2008 \cite{24-mstw}, and NNPDF2.3 \cite{21-nnpdf} PDFs. In the
calculations of the uncertainties from different PDF sets, the five
files abm11$\_$5n$\_$nnlo.LHgrid, CT10nnlo.LHgrid,
HERAPDF15NNLO$\_$EIG.LHgrid, MSTW2008nnlo68cl.LHgrid, and
NNPDF23$\_$nnlo$\_$as$\_$0119.LHgrid are adopted.

\par
Besides the differences between the various PDF sets, there are
experimental uncertainties associated with the experimental data
used to build the fit. The Hessian method is adopted by the ABM,
CT10, HERA, and MSTW collaborations to estimate the PDF experimental
uncertainty \cite{24-mstw}. In this method, additional sets next to the
best-fit PDF to account for the experimental uncertainties in the data
are used to build the distribution functions. The NNPDF collaboration
uses an alternative method to build the additional sets based on
Monte Carlo replicas \cite{25-pdf4lhc}.

\par
In addition to the PDF experimental uncertainty, there is also an
uncertainty due to the errors on the value of $\alpha_s$. The value
of the strong coupling constant $\alpha_s(M_Z^2)$ is obtained by
fitting the experimental data together with the parametric
uncertainties related to the experimental errors. That is the PDF
$\alpha_s$ uncertainty due to the variation of the $\alpha_s$ value,
which is sizeable and should be included in the total uncertainty.
We evaluate the $68\%$ C.L. $\alpha_s$ errors by taking
$\Delta\alpha_s=\pm 0.0012$ \cite{25-pdf4lhc, 26-pdf4lhc}.

\par
In Table \ref{tab2}, we list the NNLO QCD corrected total cross
sections together with the PDF experimental uncertainty and
$\alpha_s$ uncertainty at $68\%$ C.L. obtained by adopting the ABM11,
CT10, HERAPDF1.5, MSTW2008, and NNPDF2.3 PDF sets, separately. We can
see that there are obvious discrepancies between the central values
by using above five PDF sets. At the $\sqrt{S}=14~{\rm TeV}$ ($33,~
100~{\rm TeV}$) hadron collider, the smallest central prediction is
obtained from the NNPDF2.3 PDF set, which is about $3.2\%$ ($3.9\%$, $3.9\%$)
smaller than the largest one predicted by adopting the ABM11 PDF
set. In case with fixed colliding energy, the second largest central
prediction is provided by the CT10 PDF set, which is about $2.2\%$
larger than the smallest central prediction. The MSTW2008 PDF set provides
the second smallest prediction at the $\sqrt{S}=14~{\rm TeV}$
($33~{\rm TeV},~100~{\rm TeV}$) hadron collider. For each figure
in this table, the first error is from PDF
experimental uncertainty, and the second error is from $\alpha_s$
uncertainty. The data group obtained by adopting ABM11 PDFs shows
about $\pm 1\%$ combined relative PDF $+ \alpha_s$ uncertainty
(i.e., PDF experimental relative uncertainty plus $\alpha_s$
relative uncertainty), and the other four data groups by adopting
CT10, HERAPDF1.5, MSTW2008, and NNPDF2.3 PDFs show about $\pm
(1.7-3.0)\%$ combined relative PDF $+ \alpha_s$ uncertainties. The
table shows clearly that the PDF experimental error is larger than
the $\alpha_s$ error. For example, in the case of $\sqrt{S}=14~{\rm
TeV}$, the MSTW2008 PDF experimental relative error is about
$(+2.4\%\diagup-1.7\%)$, while the $\alpha_s$ error is only
$(+0.05\%\diagup-0.05\%)$. However, the predictions of the total cross
section with fixed $\sqrt{S}$ by adopting the CT10, HERAPDF1.5,
MSTW2008, and NNPDF2.3 PDF sets are in agreement within the
deviations from combined PDF experimental and $\alpha_s$
uncertainties at $68\%$ C.L., except those obtained by using ABM11
PDFs. The total error of the total cross section can be figured out
by adding linearly the scale and PDF $+ \alpha_s$ uncertainties.
According to the data in Table \ref{tab1} and Table \ref{tab2} which
are obtained by using the MSTW2008 ($68\%$ C.L.) PDFs, we can get the
total relative errors of the total cross section as
$(+4.7\%\diagup-1.8\%)$, $(+5.1\%\diagup-2.2\%)$, and
$(+6.9\%\diagup-2.8\%)$ for the $\sqrt{S}=14~{\rm TeV}$, $33$ and
$100~{\rm TeV}$ hadron colliders, separately.
\begin{table}
\begin{center}
\begin{tabular}{|c|c|c|c|}
\hline  PDF sets  & $\sqrt{S}=14~TeV$ [fb] & $\sqrt{S}=33~TeV$ [fb] & $\sqrt{S}=100~TeV$ [fb] \\
\hline  ABM11     & $2.048^{+0.020+0.003}_{-0.014-0.004}$ & $12.475^{+0.113+0.038}_{-0.071-0.038}$ &
$83.20^{+0.68+0.259}_{-0.63-0.234}$ \\
        CT10      & $2.023^{+0.039+0.001}_{-0.037-0.001}$ & $12.255^{+0.210+0.022}_{-0.201-0.013}$ &
$81.74^{+1.28+0.255}_{-1.48-0.288}$ \\
        HERA1.5   & $2.013^{+0.051+0.004}_{-0.044-0.006}$ & $12.136^{+0.269+0.022}_{-0.232-0.030}$ &
$80.45^{+1.27+0.145}_{-1.41-0.159}$ \\
        MSTW2008  & $1.986^{+0.047+0.001}_{-0.034-0.001}$ & $12.041^{+0.240+0.018}_{-0.184-0.025}$ &
$80.05^{+1.33+0.246}_{-1.17-0.309}$ \\
        NNPDF2.3  & $1.981^{+0.044+0.002}_{-0.045-0.007}$ & $11.987^{+0.221+0.047}_{-0.249-0.080}$ &
$79.97^{+1.38+0.487}_{-1.67-0.749}$ \\
\hline
\end{tabular}
\caption{\label{tab2}  The NNLO QCD corrected total cross sections
together with the $68\%$ C.L. PDF experimental and PDF $\alpha_s$
uncertainties obtained by adopting the ABM11, CT10, HERAPDF1.5,
MSTW2008 ($68\%$ C.L.) and NNPDF2.3 PDFs at the $\sqrt{S}=14~{\rm TeV}$, $33$
and $100~{\rm TeV}$ hadron colliders. For each result, the first error
is from the PDF experimental uncertainty, and the second error is due
to the $\alpha_s$ uncertainty. }
\end{center}
\end{table}

\par
\subsection{Trilinear Higgs self-coupling }
\par
The SM Higgs potential can be written as
\begin{eqnarray}\label{HiggsPotential}
V(\Phi) = \lambda (\Phi^{\dagger} \Phi)^2 - \frac{1}{2} M_H^2
\Phi^{\dagger}\Phi,
\end{eqnarray}
where
\begin{eqnarray}
\Phi = {G^+ \choose \frac{v+H+iG}{\sqrt{2}}},
\end{eqnarray}
and $G^+$ and $G$ are charged and neutral Goldstone bosons. We can
rewrite the Higgs potential Eq.(\ref{HiggsPotential}) in terms of
Higgs field $H$ as
\begin{eqnarray}
V(H) = \frac{1}{2} M_H^2 H^2 + \lambda v H^3 + \frac{\lambda}{4}
H^4,
\end{eqnarray}
where the Higgs vacuum expectation value $v =
\sqrt{\frac{M_H^2}{2\lambda}}$. Then, the SM trilinear Higgs
self-coupling $\lambda_{HHH}^{SM} = \lambda v = \frac{M_H^2}{2v}$.
\begin{figure*}
\begin{center}
\includegraphics[scale=0.65]{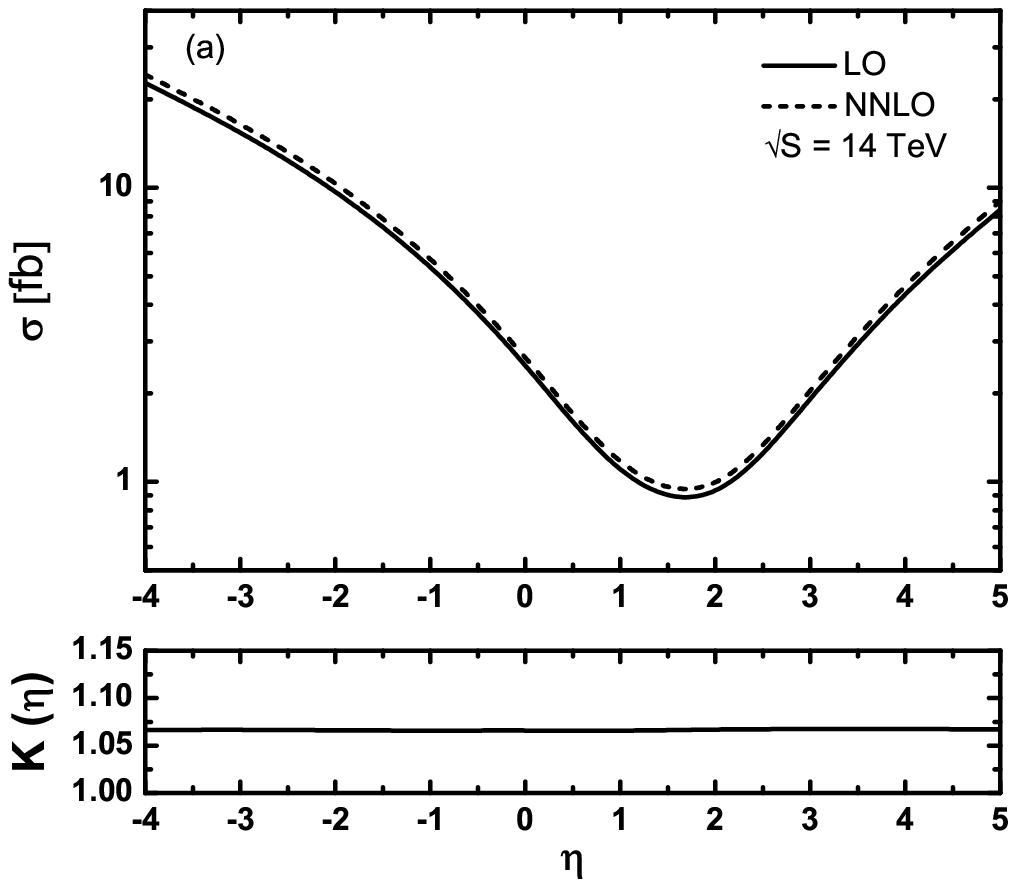}
\includegraphics[scale=0.65]{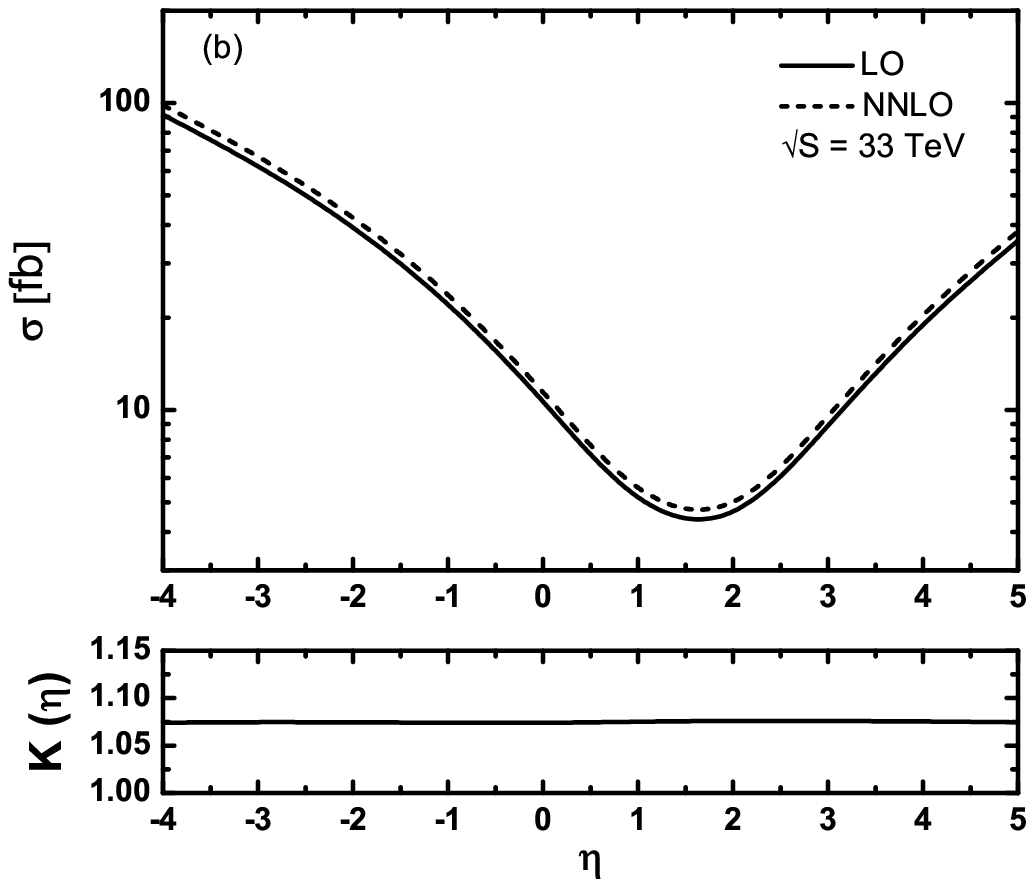}
\includegraphics[scale=0.65]{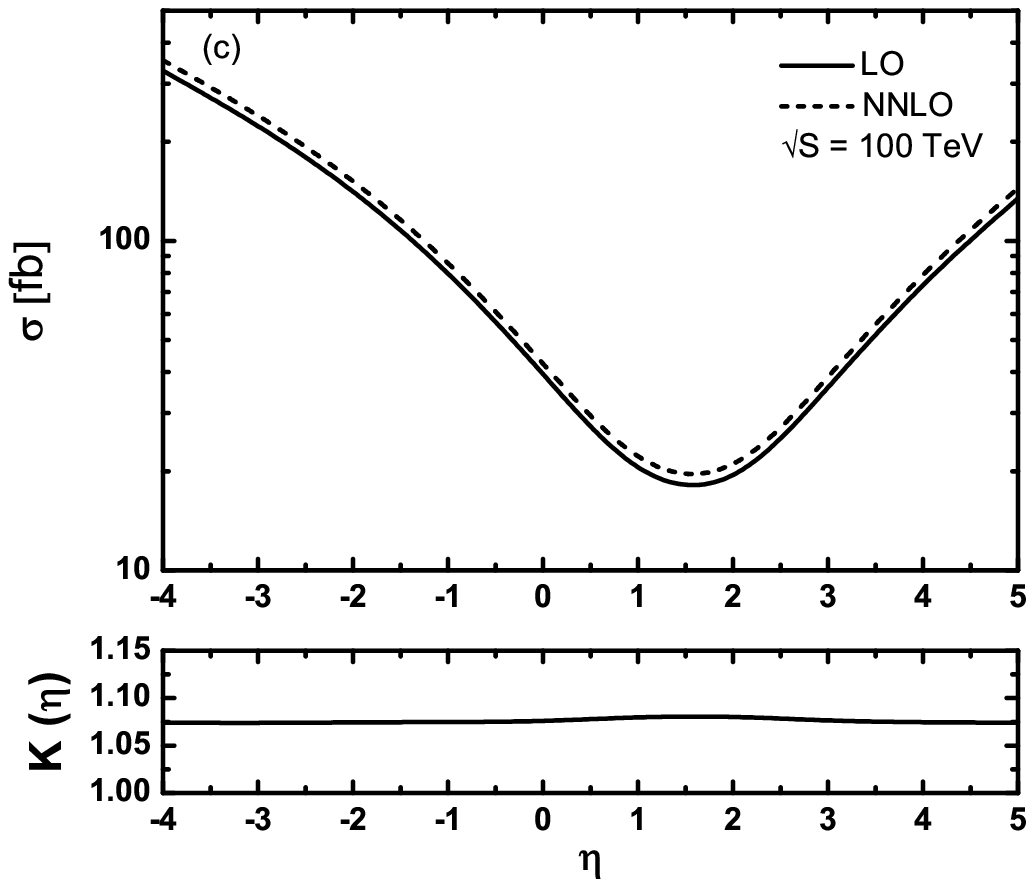}
\caption{\label{fig5}The LO and NNLO QCD corrected total cross
sections and the corresponding $K$ factors as functions of $\eta$
by using the MSTW2008 (68$\%$ C.L.) PDFs where
$\eta=\lambda_{HHH}/\lambda_{HHH}^{SM}$. (a) $\sqrt{S}=14~{\rm
TeV}$. (b) $\sqrt{S}=33~{\rm TeV}$. (c)$\sqrt{S}=100~{\rm TeV}$. }
\end{center}
\end{figure*}

\par
We deviate the trilinear Higgs self-coupling from its SM value by
substituting $\lambda_{HHH}(= \eta \lambda_{HHH}^{SM})$ for
$\lambda_{HHH}^{SM}$ to study the sensitivity of the LO and NNLO QCD
corrected total cross sections to the trilinear Higgs self-coupling
strength by using the MSTW2008 (68$\%$ C.L.) PDFs.
We find that both the LO and NNLO QCD corrected total cross sections
are strongly dependent on the parameter $\eta$, as exemplified in
Figs.\ref{fig5}(a,b,c). There Figs.\ref{fig5}(a), (b), and (c) are for
the VBF Higgs pair production at the $14$, $33$, and $100~{\rm TeV}$
hadron colliders, respectively. The corresponding
$K$ factors are shown in the lower plots of Figs.\ref{fig5}(a,b,c).
We see from the figures that the $K$ factors are stable with the
variations of parameter $\eta$.

\par
\subsection{Kinematic distributions}
\par
The signal of the VBF Higgs pair production is similar to the VBF
single Higgs production. It involves two energetic forward/backward
jets associated with two central Higgs bosons
\cite{9-vbfsig,10-vbfsig}. This character plays an important role in
discriminating the signal from the heavy QCD background. Since a
precision study of the kinematic distributions of the final
particles for the VBF Higgs pair production process is very helpful
in the theoretical and experimental analyses, we provide the NNLO
QCD corrected distributions of the transverse momenta ($p_T$) and
the rapidities ($y$) for the final Higgs bosons, as well as the
invariant mass ($m_{HH}$) and the azimuthal angle separations
($\Delta \phi$) of the final Higgs bosons. In the following, we call
the Higgs boson with relatively larger transverse momentum among the
final two Higgs bosons, i.e., $p_T^{H_1}>p_T^{H_2}$, as the first
Higgs boson $H_1$ and the other Higgs boson as the second Higgs boson
$H_2$.

\par
\begin{figure*}
\begin{center}
\includegraphics[scale=0.65]{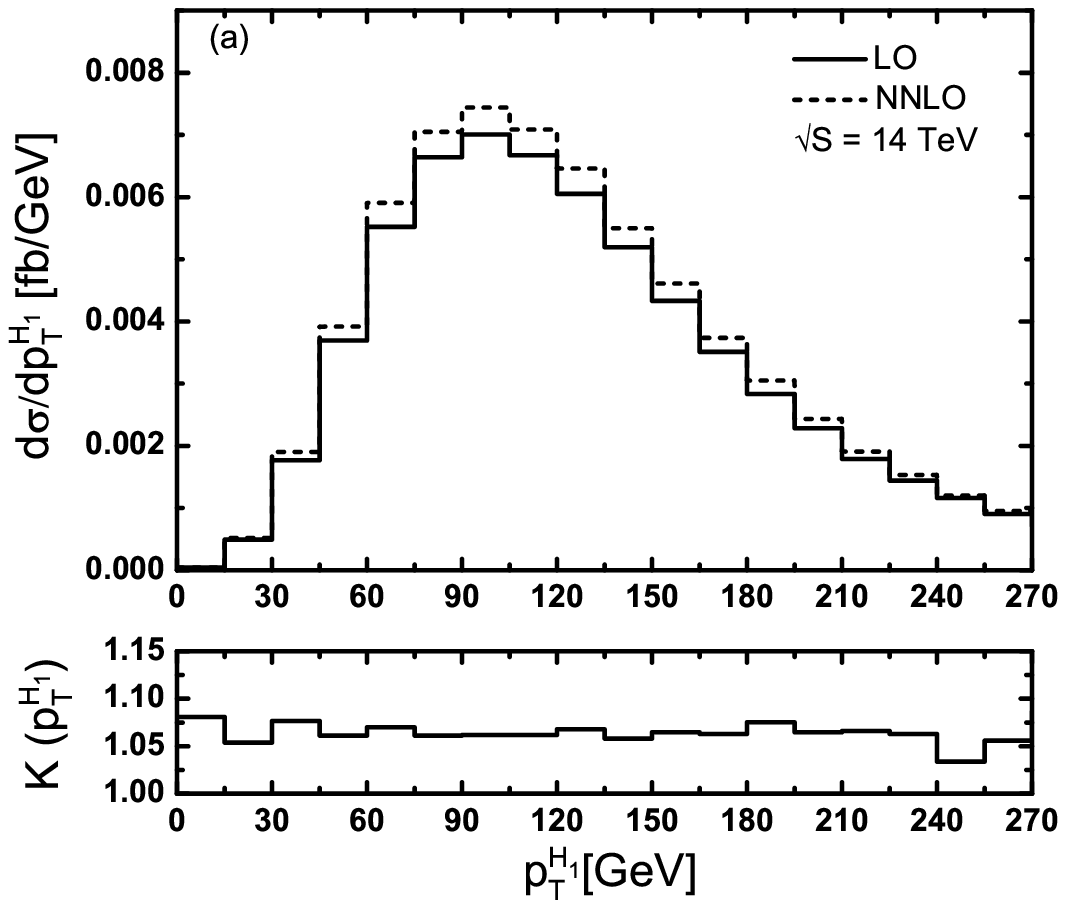}
\includegraphics[scale=0.65]{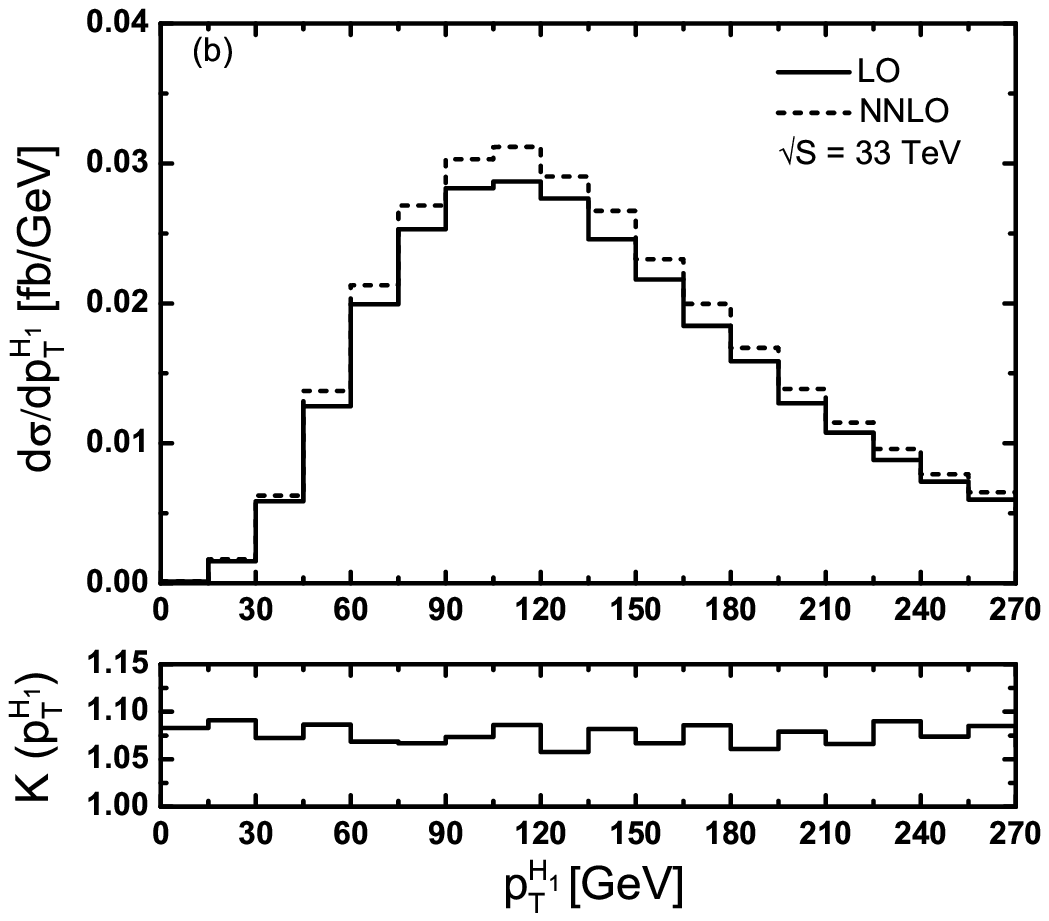}
\includegraphics[scale=0.65]{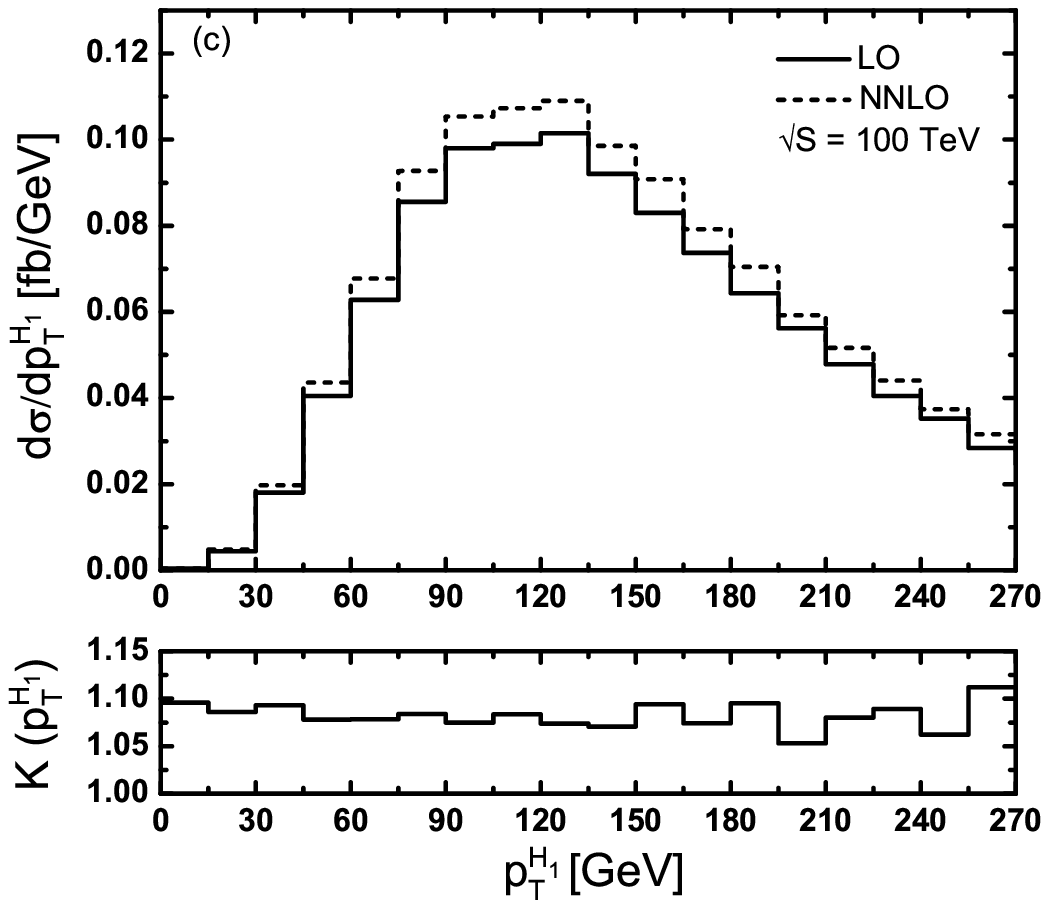}
\caption{\label{fig8} The LO and NNLO QCD corrected transverse
momentum distributions and the corresponding $K$ factors of the
first Higgs boson ($p_T^{H_1}$) for the VBF Higgs pair production
process by using the MSTW2008 (68$\%$ C.L.) PDFs.
(a) $\sqrt{S}=14~{\rm TeV}$. (b) $\sqrt{S}=33~{\rm TeV}$.
(c) $\sqrt{S}=100~{\rm TeV}$. }
\end{center}
\end{figure*}
\begin{figure*}
\begin{center}
\includegraphics[scale=0.65]{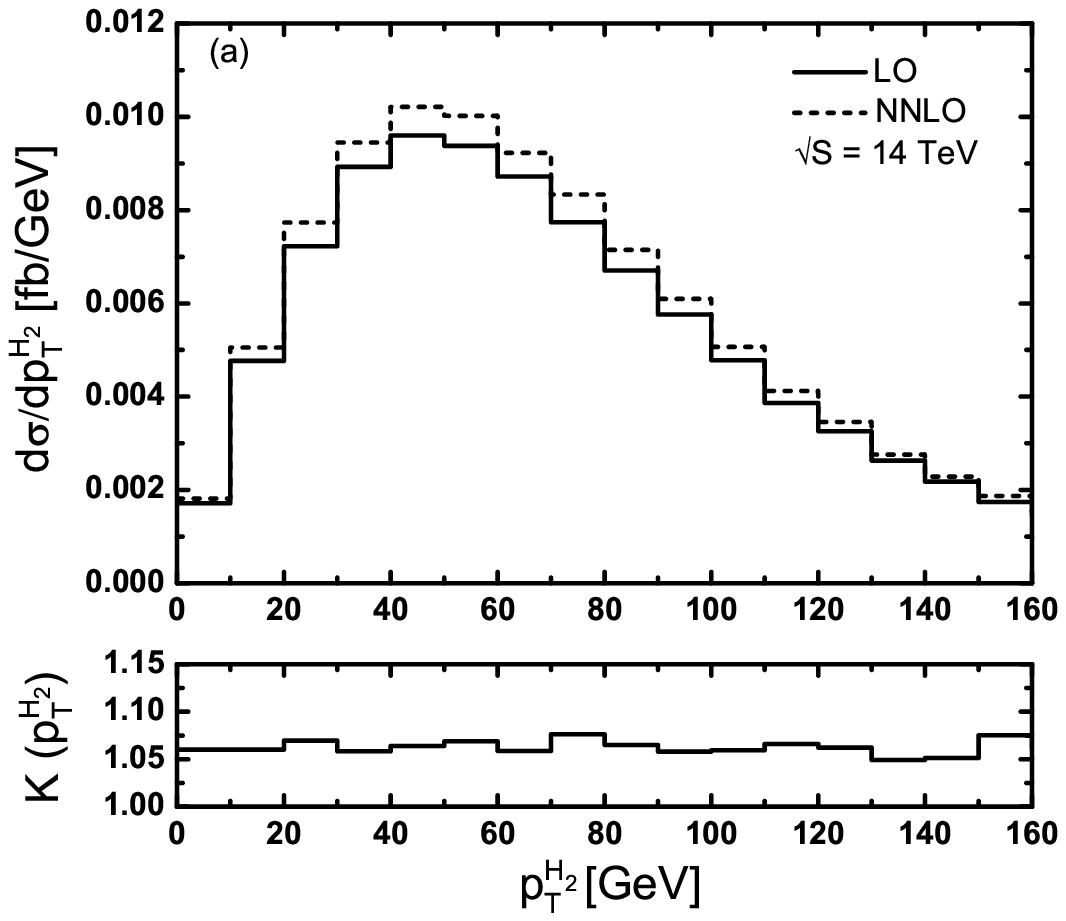}
\includegraphics[scale=0.65]{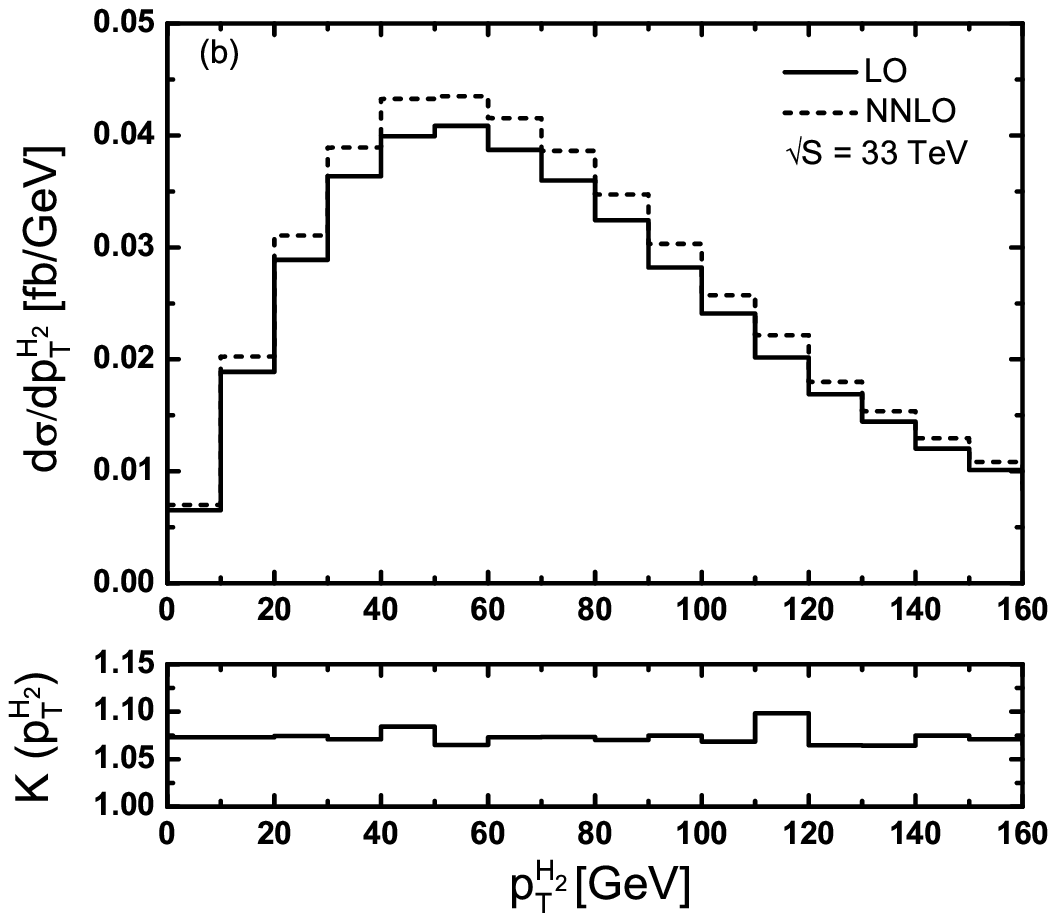}
\includegraphics[scale=0.65]{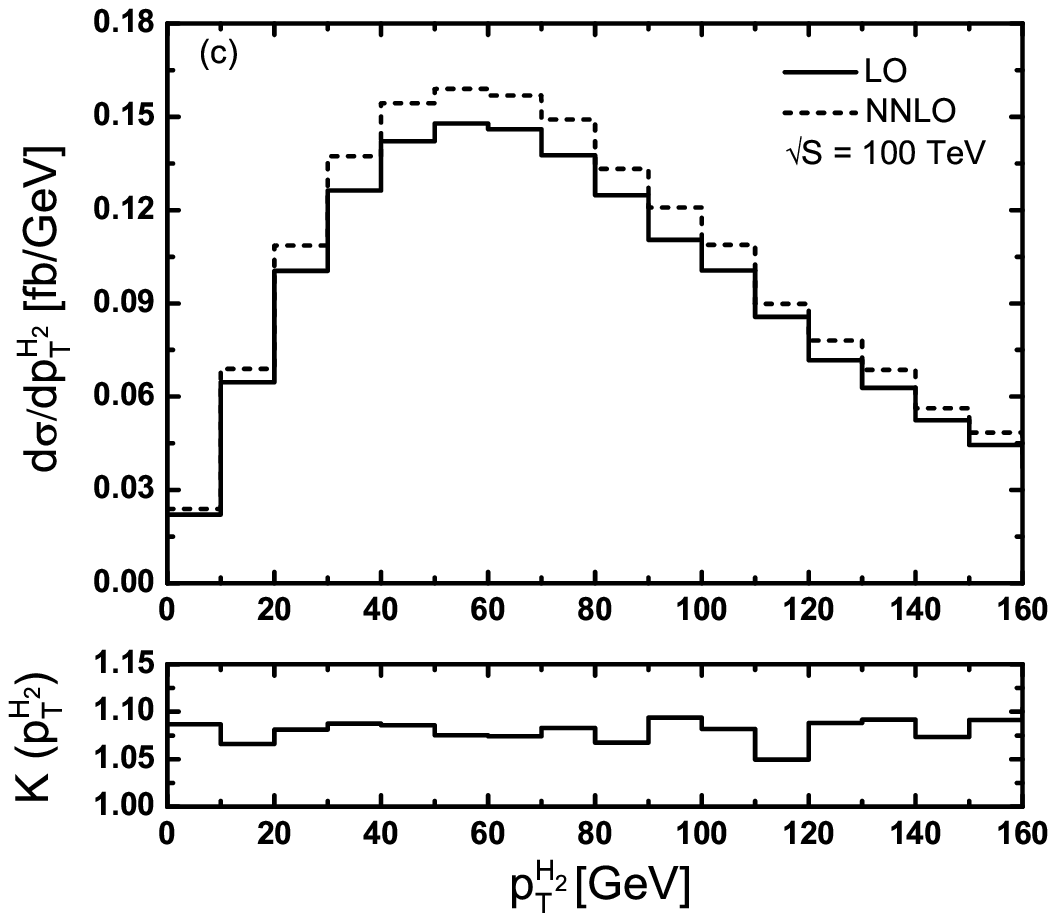}
\caption{\label{fig9} The LO and NNLO QCD corrected transverse
momentum distributions and the corresponding $K$ factors of the
second Higgs boson ($p_T^{H_2}$) for the VBF Higgs pair production
process by using the MSTW2008 (68$\%$ C.L.) PDFs.
(a) $\sqrt{S}=14~{\rm TeV}$. (b) $\sqrt{S}=33~{\rm TeV}$.
(c) $\sqrt{S}=100~{\rm TeV}$. }
\end{center}
\end{figure*}

By adopting the structure function approach, we can retain the
differential information of final Higgs bosons up to QCD NNLO but
obtain a rigorous description of final jets only at LO \cite{Zaro}.
Therefore, we only provide the kinematic distributions for final
Higgs bosons. The LO and NNLO QCD corrected transverse momentum
distributions ($\frac{d\sigma_{LO}}{dp_T}$,
$\frac{d\sigma_{NNLO}}{dp_T}$) and the corresponding $K$ factors for
the first Higgs boson at the $14$, $33$ and $100~{\rm TeV}$
hadron colliders are shown in Figs.\ref{fig8}(a,b,c),
separately. Figs. \ref{fig9}(a,b,c) demonstrate
the LO and NNLO QCD corrected $p_T$ distributions and the
corresponding $K$ factors of the second Higgs at the
$\sqrt{S}=14~{\rm TeV}$, $33$ and $100~{\rm TeV}$ hadron
colliders, separately. We see from these six figures that the
$K$ factors of the NNLO QCD corrections are less than $1.10$ in the
plotted $p_T$ range, and the $p_T$ distributions of the first Higgs
reach their maxima at the positions of $p_T\sim 90$,
$p_T\sim 100$, and $p_T\sim 100~{\rm GeV}$ at the $14$,
$33$, and $100~{\rm TeV}$ hadron colliders,
respectively, while the transverse momentum distributions of the
second Higgs boson arrive their maxima at the position of $p_T\sim
50~{\rm GeV}$ at these three hadron colliders.

\par
\begin{figure*}
\begin{center}
\includegraphics[scale=0.65]{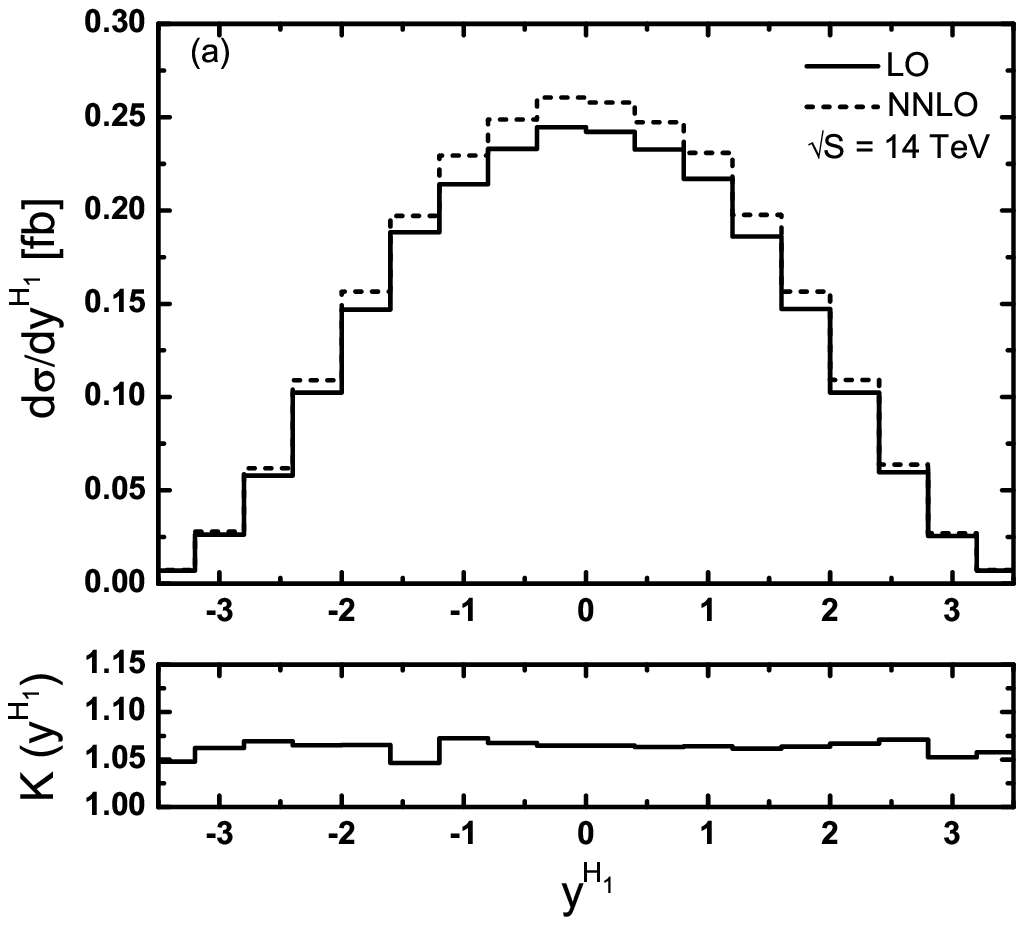}
\includegraphics[scale=0.65]{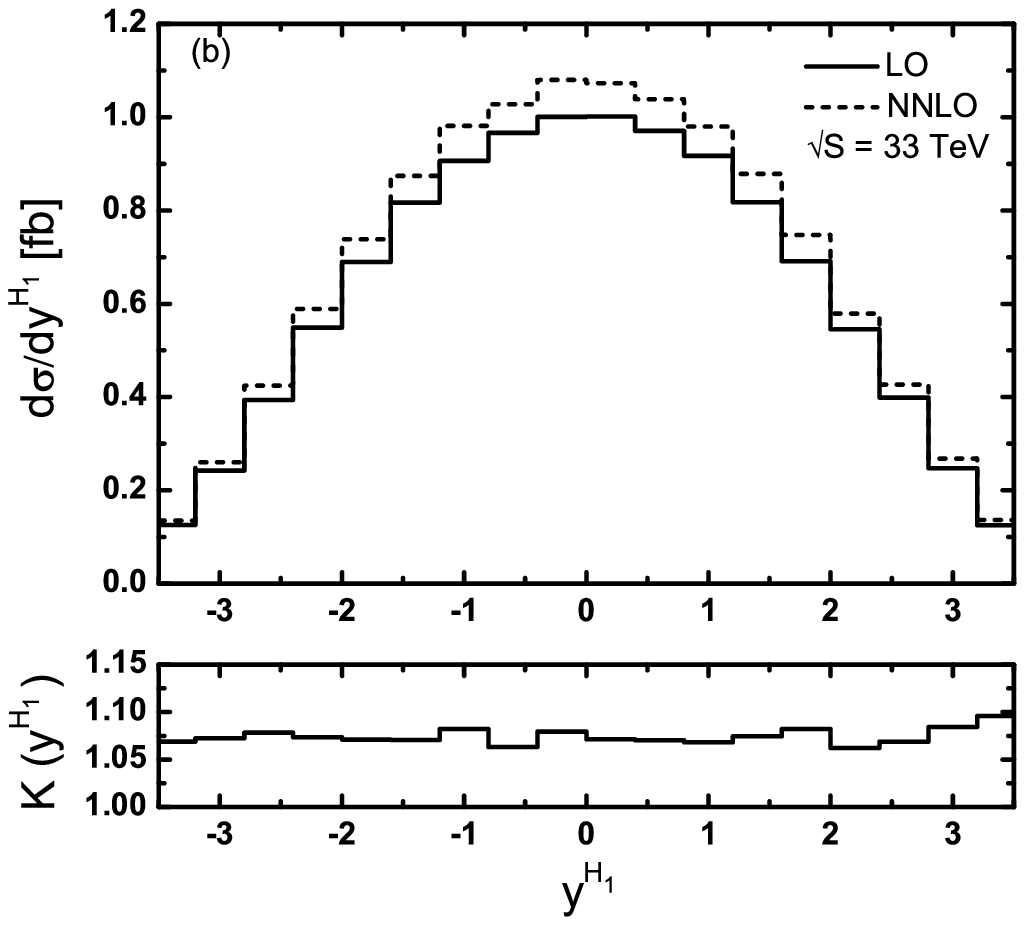}
\includegraphics[scale=0.65]{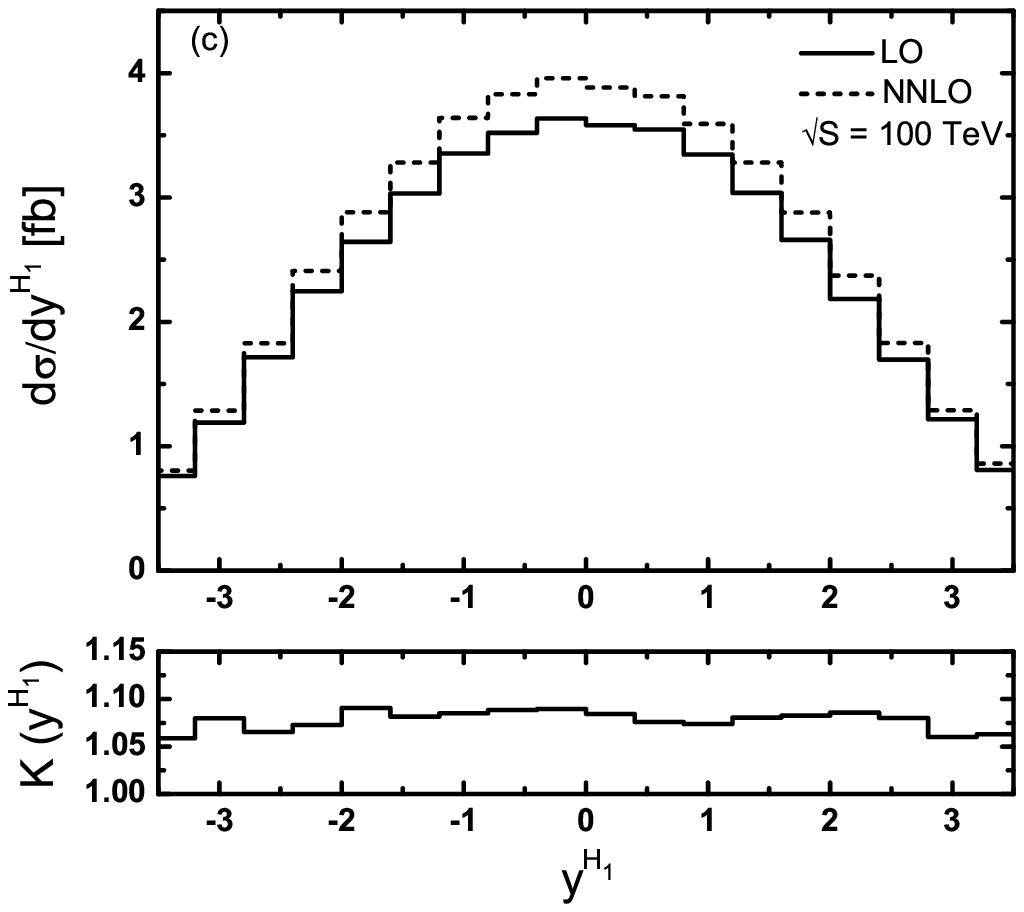}
\caption{\label{fig12} The LO and NNLO QCD corrected rapidity
distributions and the corresponding $K$ factors of the first Higgs
($y^{H_1}$) for the VBF Higgs pair production process by using the
MSTW2008 ($68\%$ C.L.) PDFs. (a) $\sqrt{S}=14~{\rm TeV}$.
(b) $\sqrt{S}=33~{\rm TeV}$. (c) $\sqrt{S}=100~{\rm TeV}$. }
\end{center}
\end{figure*}
\begin{figure*}
\begin{center}
\includegraphics[scale=0.65]{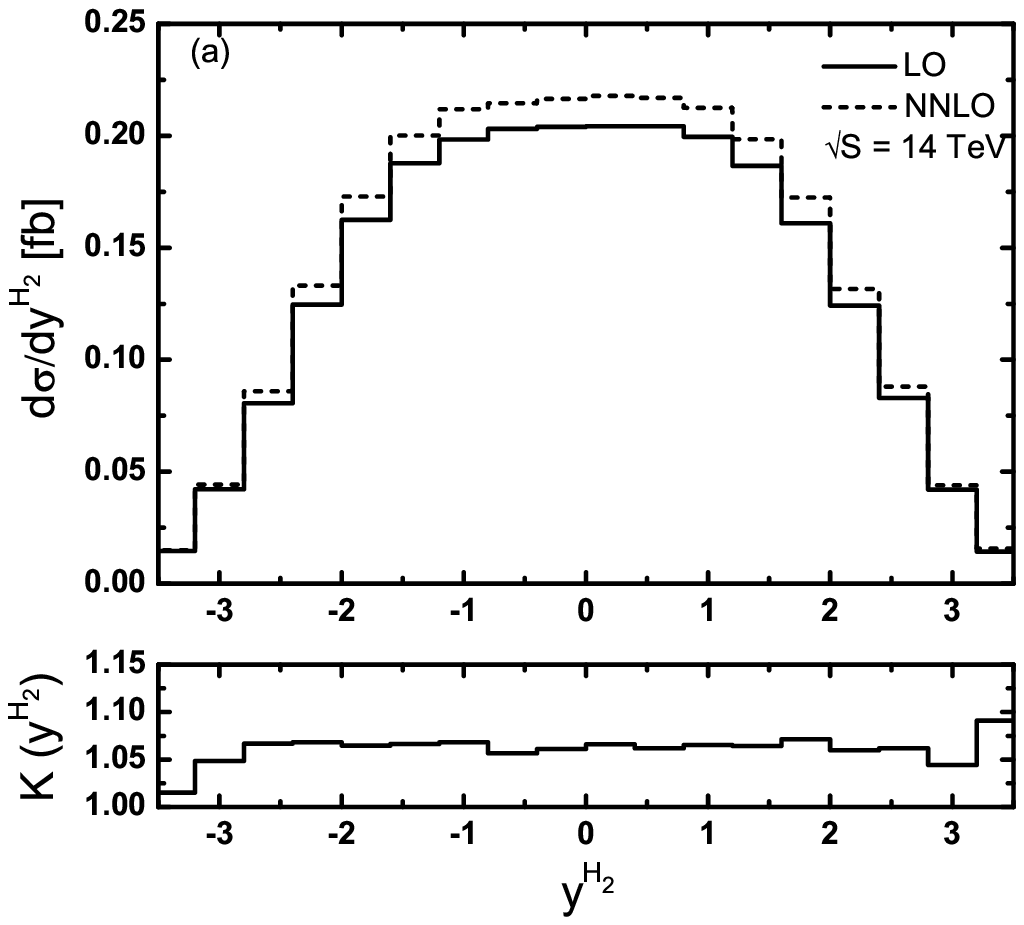}
\includegraphics[scale=0.65]{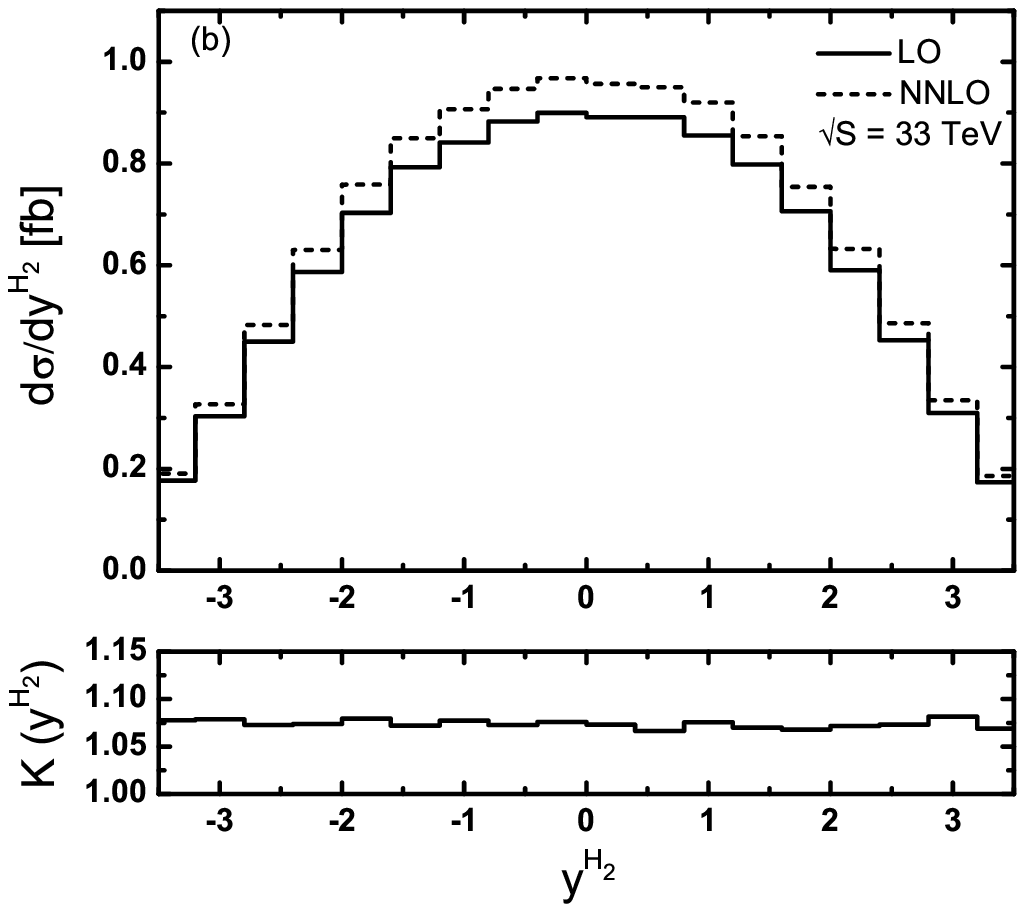}
\includegraphics[scale=0.65]{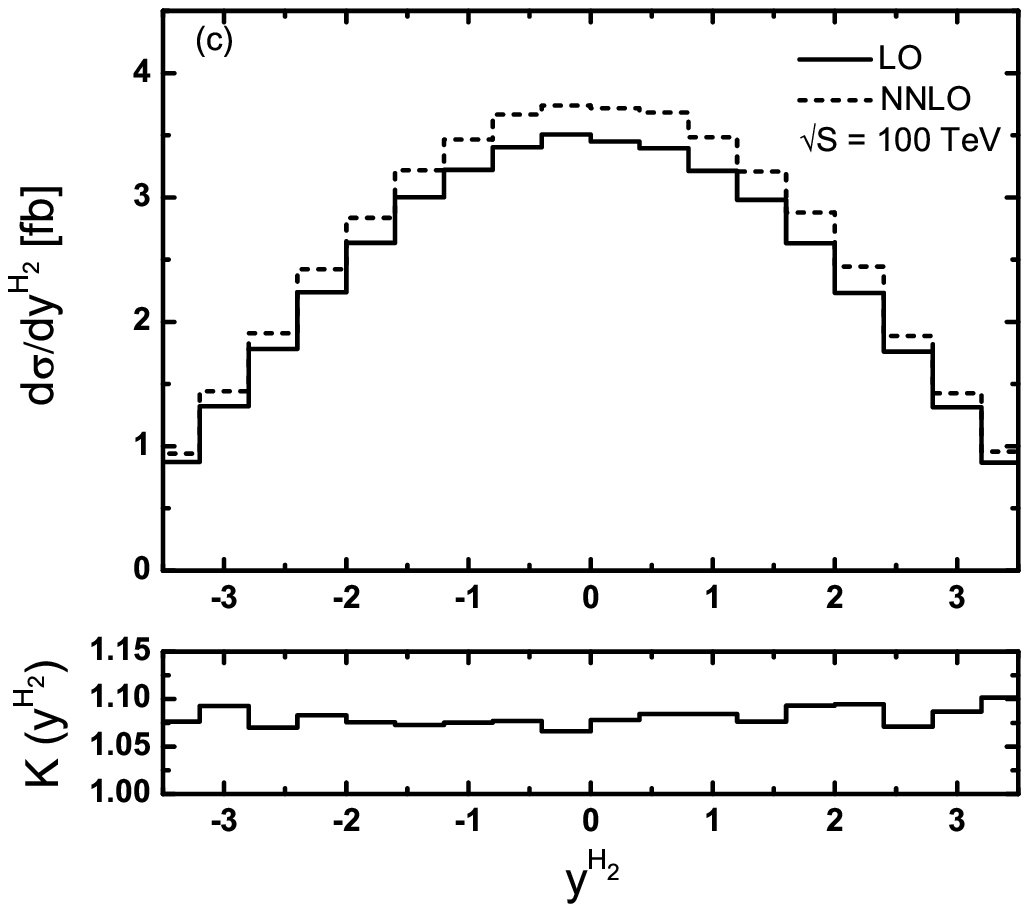}
\caption{\label{fig13} The LO and NNLO QCD corrected rapidity
distributions and the corresponding $K$ factors of the second Higgs
($y^{H_2}$) for the VBF Higgs pair production process by using the
MSTW2008 ($68\%$ C.L.) PDFs. (a) $\sqrt{S}=14~{\rm TeV}$.
(b) $\sqrt{S}=33~{\rm TeV}$. (c) $\sqrt{S}=100~{\rm TeV}$. }
\end{center}
\end{figure*}

The LO and NNLO QCD corrected rapidity distributions and the
corresponding $K$ factors of the first Higgs and second Higgs boson
by using the MSTW2008 ($68\%$ C.L.) PDFs are shown in
Figs.\ref{fig12}(a,b,c) and Figs.\ref{fig13}(a,b,c), respectively.
Figures \ref{fig12}(a) and \ref{fig13}(a) are for the $14~{\rm
TeV}$ LHC, Figs. \ref{fig12}(b) and \ref{fig13}(b) are for the
$33~{\rm TeV}$ hadron collider, and Figs. \ref{fig12}(c) and
\ref{fig13}(c) are for the $100~{\rm TeV}$ hadron collider,
separately. We can see that the two final Higgs bosons prefer to be
produced in the central rapidity region with dozens of ${\rm GeV}$
transverse momentum (see Figs. \ref{fig8} and \ref{fig9}
together). These characteristic distributions play an important role
to discriminate the signal from the very heavy QCD background
\cite{9-vbfsig,10-vbfsig}.

\par
\begin{figure*}
\begin{center}
\includegraphics[scale=0.65]{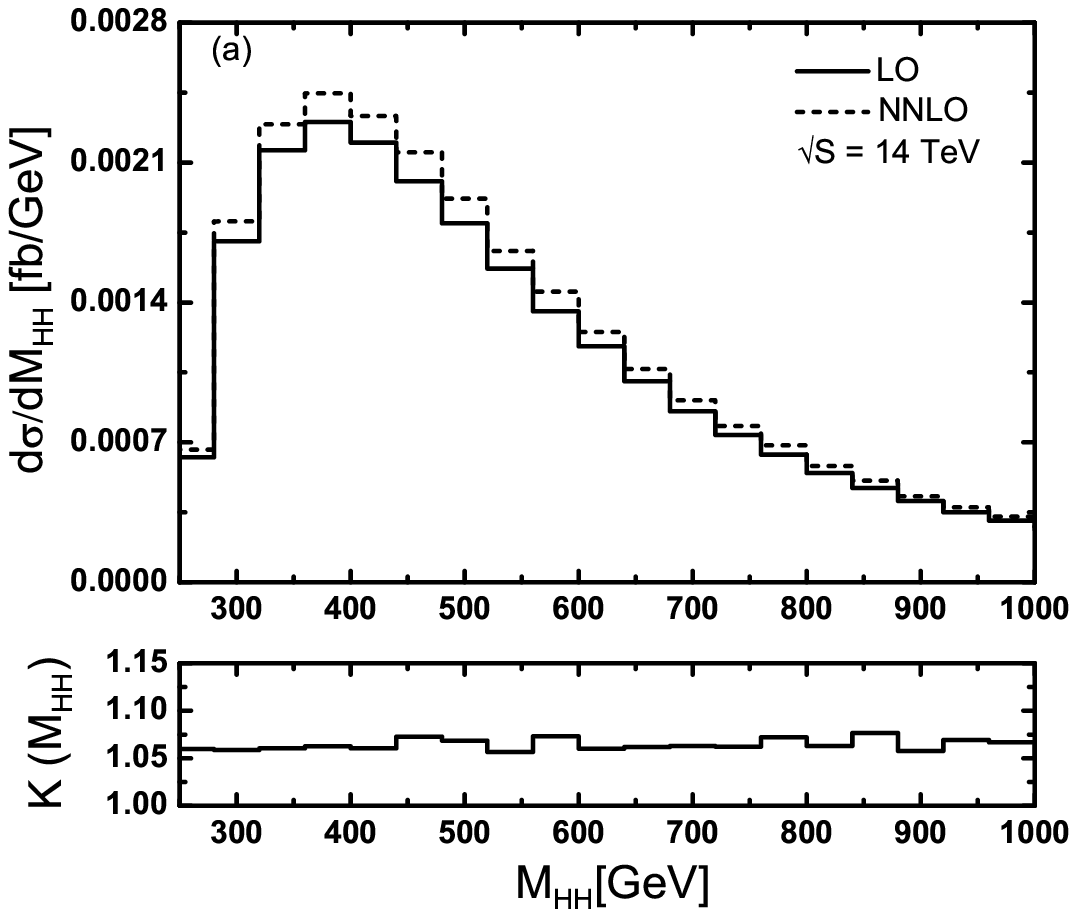}
\includegraphics[scale=0.65]{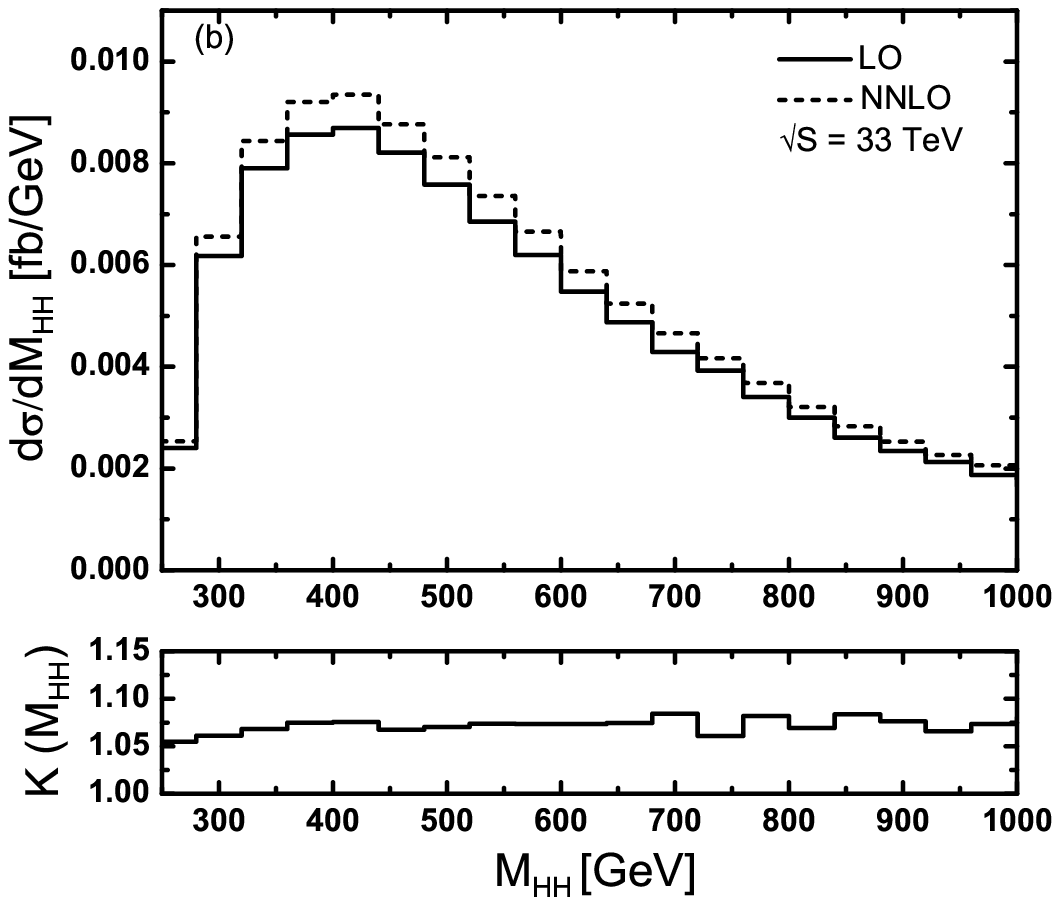}
\includegraphics[scale=0.65]{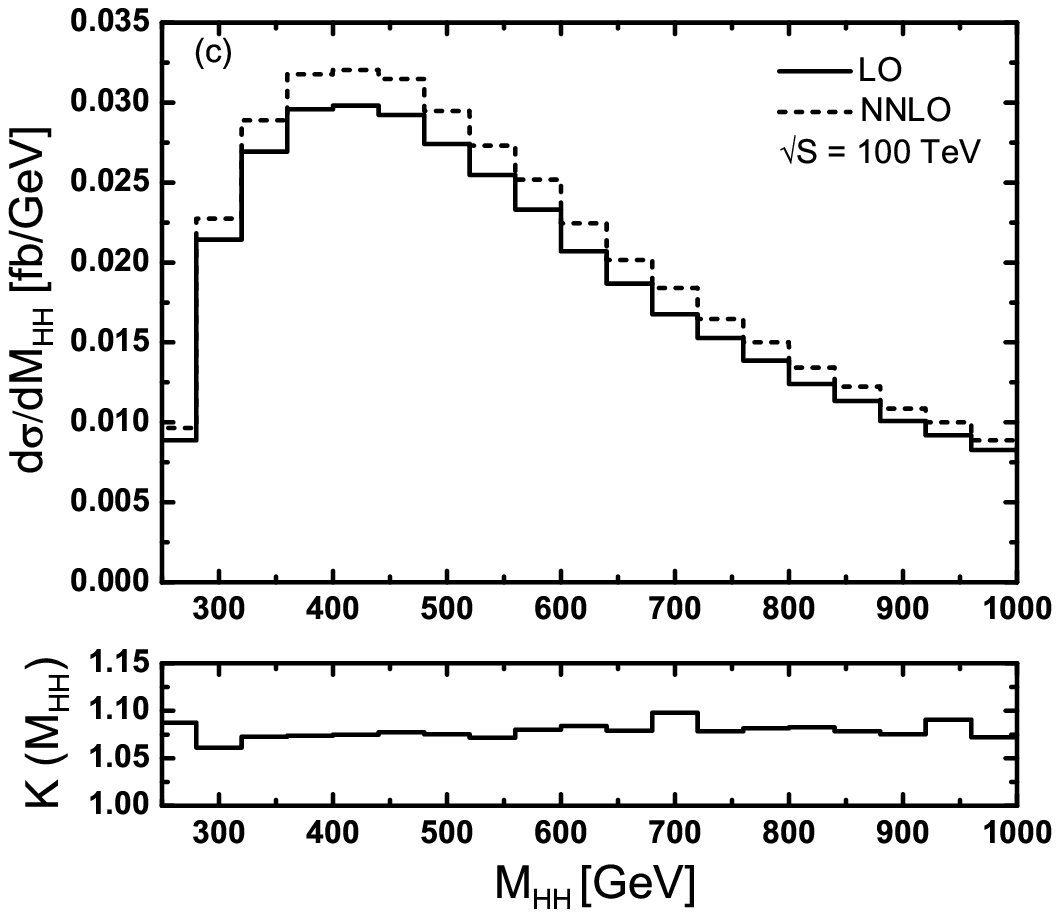}
\caption{\label{fig14} The LO and NNLO QCD corrected distributions
of the Higgs pair invariant mass ($M_{HH}$) and the corresponding
$K$ factors for the VBF Higgs pair production process
by using the MSTW2008 (68$\%$ C.L.) PDFs. (a) $\sqrt{S}=14~{\rm TeV}$.
(b) $\sqrt{S}=33~{\rm TeV}$. (c) $\sqrt{S}=100~{\rm TeV}$. }
\end{center}
\end{figure*}

In Figs. \ref{fig14}(a,b,c) we present the LO and NNLO QCD
corrected distributions of the invariant mass of final Higgs boson
pair ($M_{HH}$) and the corresponding $K$ factors by using the MSTW2008
($68\%$ C.L.) PDFs at the $14$, $33$, and $100~{\rm TeV}$ hadron colliders,
separately. From these figures, we see that for the VBF Higgs pair
production at the hadron colliders the $M_{HH}$ distributions are
mostly concentrated in the vicinity of $M_{HH} \sim 400~{\rm GeV}$
and then decrease slowly with the increment of $M_{HH}$. The
$K$ factor varies from $1.05$ to $1.10$ in the plotted invariant
mass range.

\par
\begin{figure*}
\begin{center}
\includegraphics[scale=0.65]{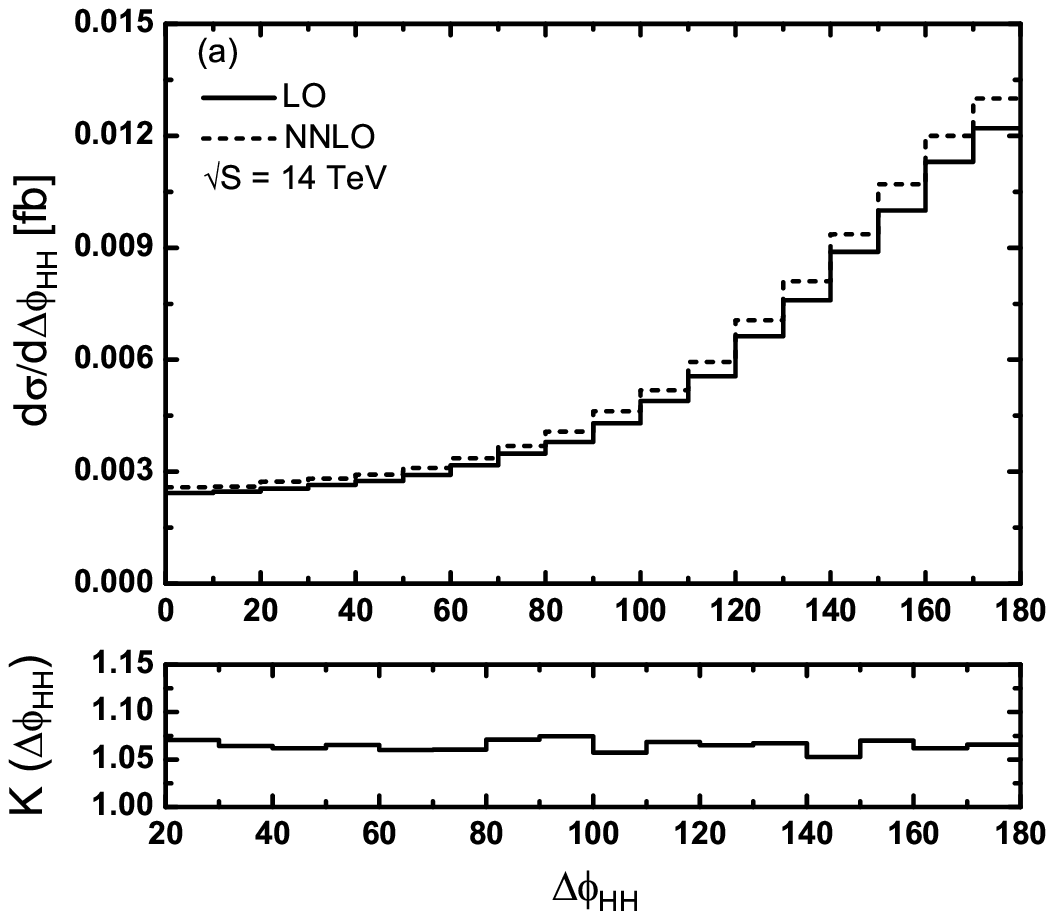}
\includegraphics[scale=0.65]{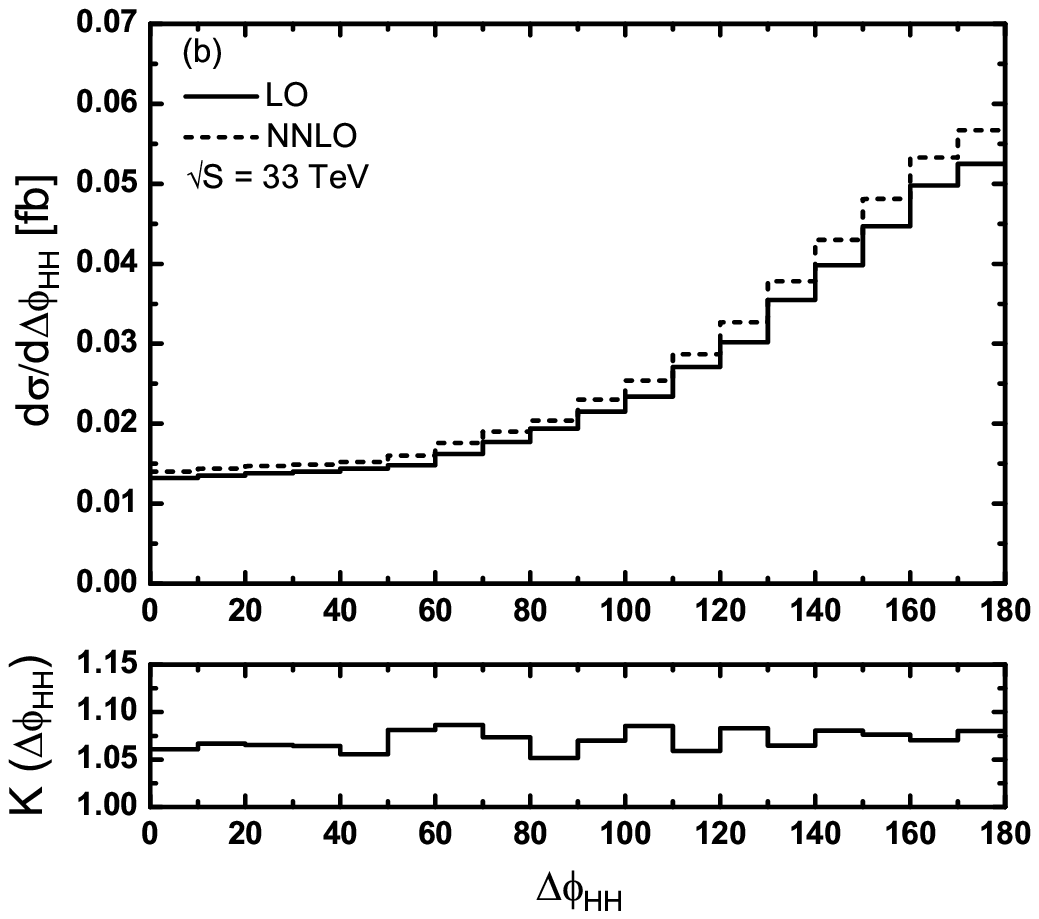}
\includegraphics[scale=0.65]{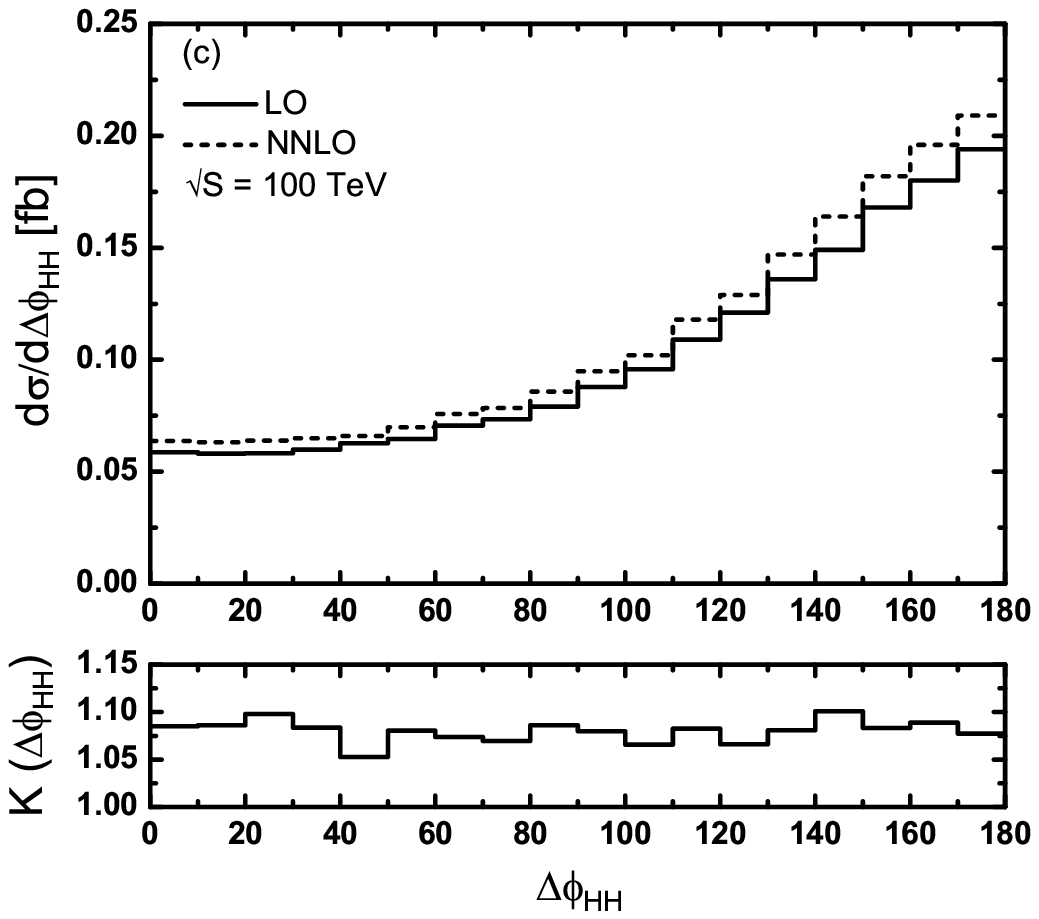}
\caption{\label{fig15} The LO and NNLO QCD corrected distributions
of the azimuthal angle separation between the final two Higgs bosons
($\Delta \phi_{HH}$) and the corresponding $K$ factors for the VBF
Higgs pair production process by using the MSTW2008 ($68\%$ C.L.) PDFs.
(a) $\sqrt{S}=14~{\rm TeV}$. (b) $\sqrt{S}=33~{\rm TeV}$.
(c) $\sqrt{S}=100~{\rm TeV}$. }
\end{center}
\end{figure*}

In Figs. \ref{fig15}(a,b,c) we present the LO and NNLO QCD
corrected distributions of the azimuthal angle separation between
the final two Higgs bosons ($\Delta \phi_{HH}$) and the
corresponding $K$ factors at the $14$, $33$,
and $100~{\rm TeV}$ hadron colliders, separately. There, we define the
azimuthal angle separation $\Delta \phi_{HH}=
|\phi_{H_1}-\phi_{H_2}|$, where $\phi_{H_1}$ and $\phi_{H_2}$ are
the azimuthal angles of the Higgs boson $H_1$ and $H_2$. The plots
show that the final two Higgs bosons prefer to be produced with
large azimuthal angle separation, and the curves for $K$ factors of
the NNLO QCD corrections are almost independent of $\Delta
\phi_{HH}$ in the plotted $\Delta \phi_{HH}$ range with the values
less than $1.10$ at the three hadron colliders.

\vskip 5mm
\section{Summary}
Probing the Higgs self-interactions is extremely significant in
understanding the EWSB mechanism. The VBF Higgs boson pair
production is an important channel in studying the trilinear Higgs
self-coupling. In this work, we calculate the NNLO QCD corrections to
the VBF SM Higgs boson pair production at the $\sqrt{S}=14$, $33$,
and $100~{\rm TeV}$ hadron colliders by using
the structure function approach. We investigate the theoretical
uncertainty from the higher-order effects by varying the
renomalization/factorization scale in the range of $[Q/4,~4Q]$ and
conclude that the total cross section at the QCD NNLO accuracy is
very stable. We also study the uncertainties from the PDFs and
$\alpha_s$ and find if we take the combined PDF and $\alpha_s$
uncertainties into account, the total cross section predictions at
the QCD NNLO by adopting the CT10, HERAPDF1.5, MSTW2008 ($68\%$ C.L.),
and NNPDF2.3 PDF sets are in good agreement. We show also the
sensitivity of the total cross section to the trilinear Higgs
self-coupling and provide the distributions of the transverse
momenta, rapidities, invariant mass, as well as the azimuthal angle
separations of the final Higgs bosons.

\vskip 5mm
\par
\noindent{\large\bf Acknowledgments:} This work was supported in
part by the National Natural Science Foundation of China (Grants.
No. 11275190, No. 11375008, and No. 11375171), and the Fundamental Research
Funds for the Central Universities (Grant. No.WK2030040044). The
numerical calculations were carried out in the Supercomputing Center
of University of Science and Technology of China.

\vskip 5mm {\bf Appendix:  Expressions for $C_{ij}$}
\par
By introducing the notations of
\begin{eqnarray}
A &=& 2 \sqrt{2} G_F \left[\frac{2
M^4_{V}}{(q_1+k_1)^2-M^2_{V}}+\frac{2 M^4_{V}}{(q_1+k_2)^2-M^2_{V}}
+\frac{6 v\lambda^{SM}_{HHH} M^2_{V}}{(k_1+k_2)^2-M^2_H}+M^2_{V}
\right], \nonumber
\\
B &=& \sqrt{2} G_F \frac{M^2_{V}}{(q_1+k_1)^2-M^2_{V}},\nonumber
\\
C &=& \sqrt{2} G_F \frac{M^2_{V}}{(q_1+k_2)^2-M^2_{V}},\nonumber
\end{eqnarray}
the coefficients $C_{ij}$ appeared in Eq.(\ref{Cij}) can be
expressed as
\begin{eqnarray}
C_{11} &=& \frac{1}{Q^2_1 Q^2_2}\Big\{A^2\Big[(q_1\cdot q_2)^2+2 Q^2_1 Q^2_2\Big]+4 A B\Big[Q^2_1 Q^2_2
(k_1\cdot k_2-M^2_H)
\nonumber\\
&&+Q^2_2(q_1\cdot k_1)(q_1\cdot k_2-q_1\cdot k_1)\Big]-4 A B (q_1\cdot q_2+q_2\cdot k_1-q_2\cdot k_2)
\nonumber\\
&&\times\Big[(q_1\cdot k_1)(q_1\cdot q_2)+(q_2\cdot k_1) Q^2_1\Big]+4 A C\Big[Q^2_1 Q^2_2(k_1\cdot k_2-M^2_H)
\nonumber\\
&&+Q^2_2(q_1\cdot k_2)(q_1\cdot k_1-q_1\cdot k_2)\Big]-4 A C (q_1\cdot q_2-q_2\cdot k_1+q_2\cdot k_2)
\nonumber\\
&&\times\Big[(q_1\cdot k_2)(q_1\cdot q_2)+(q_2\cdot k_2) Q^2_1\Big]+4B^2(q_1\cdot q_2+q_2\cdot k_1-q_2\cdot k_2)^2
\nonumber\\
&&\times\Big[M^2_HQ^2_1+(q_1\cdot k_1)^2\Big]-4B^2 Q^2_2\Big[M^2_HQ^2_1+(q_1\cdot k_1)^2\Big]
\nonumber\\
&&\times\Big[2 (k_1\cdot k_2-M^2_H-q_1\cdot k_1+q_1\cdot k_2)+Q^2_1\Big]
+8BC \Big[(k_1\cdot k_2)Q^2_1+(q_1\cdot k_1)(q_1\cdot k_2)\Big]
\nonumber\\
&&\times\Big[Q^2_2(2 k_1\cdot k_2-2 M^2_H-Q^2_1)+(q_1\cdot q_2)^2-(q_2\cdot k_1-q_2\cdot k_2)^2\Big]
\nonumber\\
&&+4C^2 (q_1\cdot q_2-q_2\cdot k_1+q_2\cdot k_2)^2\Big[M^2_H Q^2_1+(q_1\cdot k_2)^2\Big]
\nonumber\\
&&-4C^2 Q^2_2 \Big[M^2_H Q^2_1+(q_1\cdot k_2)^2\Big] \Big[2 (k_1\cdot k_2-M^2_H+q_1\cdot k_1-q_1\cdot k_2)
+Q^2_1\Big]\Big\},
\\
C_{12} &=& \frac{1}{(P_2\cdot q_2) Q^2_1 Q^4_2}\Big\{-A^2 \Big[(P_2\cdot q_1)^2 Q^4_2+2 (P_2\cdot q_1)
(P_2\cdot q_2)(q_1\cdot q_2) Q^2_2
\nonumber\\
&&+(P_2\cdot q_2)^2(q_1\cdot q_2)^2+Q^2_1 Q^2_2(P_2\cdot q_2)^2\Big]+4AB\Big[Q^2_2 (P_2\cdot k_1
-P_2\cdot k_2+P_2\cdot q_1)
\nonumber\\
&&+(P_2\cdot q_2) (q_1\cdot q_2+q_2\cdot k_1-q_2\cdot k_2)\Big]\Big[(P_2\cdot k_1)Q^2_1 Q^2_2
\nonumber\\
&&+(P_2\cdot q_1)(q_1\cdot k_1)Q^2_2+(P_2\cdot q_2) (q_1\cdot k_1) (q_1\cdot q_2)
+(P_2\cdot q_2) (q_2\cdot k_1) Q^2_1\Big]
\nonumber\\
&&+4AC \Big[Q^2_2 (P_2\cdot k_2-P_2\cdot k_1+P_2\cdot q_1)+(P_2\cdot q_2)(q_1\cdot q_2-q_2\cdot k_1+q_2\cdot k_2)\Big]
\nonumber\\
&&\times\Big[(P_2\cdot k_2)Q^2_1 Q^2_2+(P_2\cdot q_1) (q_1\cdot k_2) Q^2_2+(P_2\cdot q_2) (q_1\cdot k_2)(q_1\cdot q_2)
\nonumber\\
&&+(P_2\cdot q_2)(q_2\cdot k_2) Q^2_1\Big]-8BC\Big[(k_1\cdot k_2)Q^2_1+(q_1\cdot k_1)(q_1\cdot k_2)\Big]
\nonumber\\
&&\times\Big[Q^2_2(P_2\cdot k_1-P_2\cdot k_2+P_2\cdot q_1)+(P_2\cdot q_2)(q_1\cdot q_2+q_2\cdot k_1-q_2\cdot k_2)\Big]
\nonumber\\
&&\times\Big[Q^2_2 (P_2\cdot k_2-P_2\cdot k_1+P_2\cdot q_1)+(P_2\cdot q_2)(q_1\cdot q_2-q_2\cdot k_1
+q_2\cdot k_2)\Big]
\nonumber\\
&&-4B\Big[M^2_H Q^2_1+(q_1\cdot k_1)^2\Big]\Big[Q^2_2 (P_2\cdot k_1-P_2\cdot k_2+P_2\cdot q_1)
\nonumber\\
&&+(P_2\cdot q_2)(q_1\cdot q_2+q_2\cdot k_1-q_2\cdot k_2)\Big]^2-4C^2 \Big[M^2_H Q^2_1+(q_1\cdot k_2)^2\Big]
\nonumber\\
&&\times\Big[Q^2_2 (P_2\cdot k_2-P_2\cdot k_1+P_2\cdot q_1)+(P_2\cdot q_2)(q_1\cdot q_2-q_2\cdot k_1
+q_2\cdot k_2)\Big]^2\Big\},
\\
C_{13} &=& 0,
\\
C_{21} &=& \frac{1}{(P_1\cdot q_1) Q^4_1 Q^2_2}\Big\{-A^2(P_1\cdot
q_1)^2 Q^2_1 Q^2_2-A^2\Big[(P_1\cdot q_1) (q_1\cdot q_2)+(P_1\cdot
q_2) Q^2_1\Big]^2
\nonumber\\
&&+4AB\Big[(P_1\cdot k_1)Q^2_1+(P_1\cdot q_1)(q_1\cdot k_1)\Big]\Big[Q^2_1Q^2_2(P_1\cdot k_1)
-Q^2_1Q^2_2(P_1\cdot k_2)
\nonumber\\
&&+Q^2_2(P_1\cdot q_1) (q_1\cdot k_1)-Q^2_2(P_1\cdot q_1) (q_1\cdot k_2)
+(q_1\cdot q_2+q_2\cdot k_1-q_2\cdot k_2)(P_1\cdot q_1) (q_1\cdot q_2)
\nonumber\\
&&+(q_1\cdot q_2+q_2\cdot k_1-q_2\cdot k_2)
(P_1\cdot q_2)Q^2_1\Big]+4AC\Big[(P_1\cdot k_2)Q^2_1+(P_1\cdot q_1) (q_1\cdot k_2)\Big]
\nonumber\\
&&\times\Big[Q^2_1Q^2_2(P_1\cdot k_2)-Q^2_1Q^2_2(P_1\cdot k_1)-Q^2_2(P_1\cdot q_1) (q_1\cdot k_1)+Q^2_2(P_1\cdot q_1)
(q_1\cdot k_2)
\nonumber\\
&&+(q_1\cdot q_2-q_2\cdot k_1+q_2\cdot k_2)(P_1\cdot q_1) (q_1\cdot q_2)+(q_1\cdot q_2-q_2\cdot k_1
+q_2\cdot k_2)(P_1\cdot q_2)Q^2_1\Big]
\nonumber\\
&&-4B^2\Big[(P_1\cdot k_1)Q^2_1+(P_1\cdot q_1) (q_1\cdot k_1)\Big]^2\Big[(q_1\cdot q_2+q_2\cdot k_1-q_2\cdot k_2)^2
\nonumber\\
&&-2 Q^2_2(k_1\cdot k_2-M^2_H-q_1\cdot k_1+q_1\cdot k_2)-Q^2_1Q^2_2\Big]+8BC\Big[(P_1\cdot k_1)Q^2_1
\nonumber\\
&&+(P_1\cdot q_1) (q_1\cdot k_1)\Big]\Big[(P_1\cdot k_2)Q^2_1+(P_1\cdot q_1) (q_1\cdot k_2)\Big]
\Big[Q^2_2 (Q^2_1-2 k_1\cdot k_2+2 M^2_H)
\nonumber\\
&&-(q_1\cdot q_2)^2+(q_2\cdot k_1-q_2\cdot k_2)^2\Big]-4C^2\Big[(P_1\cdot k_2)Q^2_1+(P_1\cdot q_1)
(q_1\cdot k_2)\Big]^2
\nonumber\\
&&\times\Big[(q_1\cdot q_2-q_2\cdot k_1+q_2\cdot k_2)^2-2Q^2_2 (k_1\cdot k_2-M^2_H+q_1\cdot k_1-q_1\cdot k_2)
-Q^2_1Q^2_2\Big]\Big\},
\\
C_{22} &=& \frac{1}{4 (P_1\cdot q_1)(P_2\cdot q_2)Q^4_1 Q^4_2}\Big\{-A \Big[2 (P_1\cdot q_1) (P_2\cdot q_1) Q^2_2
+2 (P_1\cdot q_1) (P_2\cdot q_2) (q_1\cdot q_2)
\nonumber\\
&&+2 (P_1\cdot q_2) (P_2\cdot q_2)Q^2_1+Q^2_1 Q^2_2 S\Big]+4 B \Big[(P_1\cdot k_1)Q^2_1
+(P_1\cdot q_1) (q_1\cdot k_1)\Big]
\nonumber\\
&&\times\Big[Q^2_2 (P_2\cdot k_1-P_2\cdot k_2+P_2\cdot q_1)+(P_2\cdot q_2)
(q_1\cdot q_2+q_2\cdot k_1-q_2\cdot k_2)\Big]
\nonumber\\
&&+4 C\Big[(P_1\cdot k_2)Q^2_1+(P_1\cdot q_1) (q_1\cdot k_2)\Big]\Big[Q^2_2 (P_2\cdot k_2-P_2\cdot k_1
+P_2\cdot q_1)
\nonumber\\
&&+(P_2\cdot q_2)(q_1\cdot q_2-q_2\cdot k_1+q_2\cdot k_2)\Big]\Big\}^2,
\\
C_{23} &=& 0,
\\
C_{31} &=& 0,
\\
C_{32} &=& 0,
\\
C_{33} &=& \frac{1}{4 (P_1\cdot q_1) (P_2\cdot q_2)}\bigg\{A^2 \Big[(q_1\cdot q_2)S-2 (P_1\cdot q_2)
(P_2\cdot q_1)\Big]-2 AB \Big\{S \Big[-(k_1\cdot k_2)(q_1\cdot q_2)               \nonumber\\
&&+M^2_H (q_1\cdot q_2)+(q_1\cdot k_1) (q_1\cdot q_2)-(q_1\cdot k_1) (q_2\cdot k_1)+(q_1\cdot k_2)
(q_2\cdot k_1)+(q_2\cdot k_1) Q^2_1\Big]                                          \nonumber\\
&&-2 \Big[M^2_H (P_1\cdot q_2)(P_2\cdot q_1)-(k_1\cdot k_2) (P_1\cdot q_2) (P_2\cdot q_1)
+(P_1\cdot k_1) (P_2\cdot k_1) (q_1\cdot q_2)                                     \nonumber\\
&&-(P_1\cdot k_1) (P_2\cdot q_1) (q_2\cdot k_1)-(P_1\cdot k_2) (P_2\cdot k_1)(q_1\cdot q_2)
+(P_1\cdot k_2) (P_2\cdot q_1) (q_2\cdot k_1)                                     \nonumber\\
&&+(P_1\cdot q_1)(P_2\cdot k_1) (q_1\cdot q_2)-(P_1\cdot q_1) (P_2\cdot q_1) (q_2\cdot k_1)
-(P_1\cdot q_2) (P_2\cdot k_1) (q_1\cdot k_1)                                     \nonumber\\
&&+(P_1\cdot q_2) (P_2\cdot k_1) (q_1\cdot k_2)+(P_1\cdot q_2) (P_2\cdot k_1) Q^2_1
+(P_1\cdot q_2)(P_2\cdot q_1) (q_1\cdot k_1)\Big]\Big\}                           \nonumber\\
&&-2 AC \Big\{S \Big[M^2_H (q_1\cdot q_2)-(k_1\cdot k_2) (q_1\cdot q_2)+(q_1\cdot k_1) (q_2\cdot k_2)
+(q_1\cdot k_2) (q_1\cdot q_2)                                                    \nonumber\\
&&-(q_1\cdot k_2) (q_2\cdot k_2)+(q_2\cdot k_2) Q^2_1\Big]
-2 \Big[M^2_H (P_1\cdot q_2) (P_2\cdot q_1)                                       \nonumber\\
&&-(k_1\cdot k_2) (P_1\cdot q_2) (P_2\cdot q_1)-(P_1\cdot k_1) (P_2\cdot k_2) (q_1\cdot q_2)
+(P_1\cdot k_1) (P_2\cdot q_1) (q_2\cdot k_2)                                     \nonumber\\
&&+(P_1\cdot k_2) (P_2\cdot k_2) (q_1\cdot q_2) -(P_1\cdot k_2) (P_2\cdot q_1) (q_2\cdot k_2)
+(P_1\cdot q_1) (P_2\cdot k_2) (q_1\cdot q_2)                                     \nonumber\\
&&-(P_1\cdot q_1) (P_2\cdot q_1) (q_2\cdot k_2)+(P_1\cdot q_2)(P_2\cdot k_2) (q_1\cdot k_1)
-(P_1\cdot q_2) (P_2\cdot k_2) (q_1\cdot k_2)                                     \nonumber\\
&&+(P_1\cdot q_2) (P_2\cdot k_2) Q^2_1+(P_1\cdot q_2) (P_2\cdot q_1)(q_1\cdot k_2)\Big]\Big\}
+8 B C \Big\{2 (k_1\cdot k_2)                                                     \nonumber\\
&&\times\Big[(P_1\cdot q_1) (q_1\cdot q_2) (P_2\cdot k_1+P_2\cdot k_2)
-(P_1\cdot q_1) (P_2\cdot q_1) (q_2\cdot k_1+q_2\cdot k_2)                        \nonumber\\
&&+Q^2_1 (P_1\cdot q_2)(P_2\cdot k_1+P_2\cdot k_2)+(P_1\cdot q_2) (P_2\cdot q_1)
(q_1\cdot k_1+q_1\cdot k_2)\Big]                                                  \nonumber\\
&&-(k_1\cdot k_2) S \Big[(q_1\cdot q_2) (q_1\cdot k_1+q_1\cdot k_2)+Q^2_1
(q_2\cdot k_1+q_2\cdot k_2)\Big]                                                  \nonumber\\
&&-2 M^2_H \Big[(P_1\cdot q_1) (q_1\cdot q_2)(P_2\cdot k_1+P_2\cdot k_2)
-(P_1\cdot q_1) (P_2\cdot q_1) (q_2\cdot k_1+q_2\cdot k_2)                        \nonumber\\
&&+(P_1\cdot q_2) Q^2_1 (P_2\cdot k_1+P_2\cdot k_2)+(P_1\cdot q_2)  (P_2\cdot q_1)
(q_1\cdot k_1+q_1\cdot k_2)\Big]                                                  \nonumber\\
&&+M^2_H S \Big[(q_1\cdot q_2) (q_1\cdot k_1+q_1\cdot k_2)+Q^2_1(q_2\cdot k_1+q_2\cdot k_2)\Big]    \nonumber\\
&&+2 \Big[Q^2_1 (P_1\cdot k_1-P_1\cdot k_2)(P_2\cdot k_2) (q_2\cdot k_1)
-Q^2_1 (P_1\cdot k_1-P_1\cdot k_2)(P_2\cdot k_1) (q_2\cdot k_2)                  \nonumber\\
&&-(P_1\cdot k_1) (P_2\cdot k_1) (q_1\cdot k_2) (q_1\cdot q_2)
+(P_1\cdot k_1) (P_2\cdot k_2) (q_1\cdot k_1) (q_1\cdot q_2)                      \nonumber\\
&&-(P_1\cdot k_1) (P_2\cdot q_1) (q_1\cdot k_1) (q_2\cdot k_2)
+(P_1\cdot k_1) (P_2\cdot q_1) (q_1\cdot k_2) (q_2\cdot k_1)                      \nonumber\\
&&+(P_1\cdot k_2) (P_2\cdot k_1) (q_1\cdot k_2) (q_1\cdot q_2)
-(P_1\cdot k_2) (P_2\cdot k_2) (q_1\cdot k_1) (q_1\cdot q_2)                      \nonumber\\
&&+(P_1\cdot k_2) (P_2\cdot q_1) (q_1\cdot k_1) (q_2\cdot k_2)
-(P_1\cdot k_2) (P_2\cdot q_1) (q_1\cdot k_2) (q_2\cdot k_1)                      \nonumber\\
&&-(P_1\cdot q_1) (P_2\cdot k_1) (q_1\cdot k_1) (q_2\cdot k_2)
+(P_1\cdot q_1) (P_2\cdot k_1) (q_1\cdot k_2) (q_2\cdot k_2)                      \nonumber\\
&&+(P_1\cdot q_1) (P_2\cdot k_2)(q_1\cdot k_1) (q_2\cdot k_1)
-(P_1\cdot q_1) (P_2\cdot k_2) (q_1\cdot k_2) (q_2\cdot k_1)                      \nonumber\\
&&-(P_1\cdot q_2)(P_2\cdot k_2) (q_1\cdot k_1)^2
+(P_1\cdot q_2)(q_1\cdot k_2)(P_2\cdot k_2) (q_1\cdot k_1)                        \nonumber\\
&&+(P_1\cdot q_2)(q_1\cdot k_1)(P_2\cdot k_1) (q_1\cdot k_2)
-(P_1\cdot q_2)(P_2\cdot k_1) (q_1\cdot k_2)^2\Big]                   \nonumber\\
&&+S \Big[(q_1\cdot k_1)-(q_1\cdot k_2)\Big]\Big[(q_1\cdot k_1) (q_2\cdot k_2)
-(q_1\cdot k_2)  (q_2\cdot k_1)\Big]\Big\}\bigg\}.
\end{eqnarray}

\end{document}